\documentclass[11pt]{article}
\usepackage{times}
\usepackage{mathptmx}
\usepackage{amsfonts}
\usepackage{amsmath}
\usepackage{graphicx}

\numberwithin{equation}{section}

\newcommand{\ie}{i.e.,\ }
\newcommand{\eg}{e.g.,\ }

\newcommand{\rmd}{\,\mathrm{d}}

\newcommand{\idmat}{\mathbb{I}}

\newcommand{\re}{\operatorname{Re}}
\newcommand{\im}{\operatorname{Im}}

\newcommand{\e}[1]{\operatorname{e}^{#1}}

\newcommand{\tik}{\tilde{k}}
\newcommand{\talpha}{\tilde{\alpha}}
\newcommand{\tbeta}{\tilde{\beta}}
\newcommand{\tchi}{\tilde{\chi}}

\newcommand{\BesK}{\operatorname{K}} 
\newcommand{\BesI}{\operatorname{I}} 

\newcommand{\UV}{\text{UV}}
\newcommand{\IR}{\text{IR}}
\newcommand{\pUV}{\phi_{\UV}}
\newcommand{\pIR}{\phi_{\IR}}

\newcommand{\sour}{\mathfrak{s}}
\newcommand{\resp}{\mathfrak{r}}

\newcommand{\prop}{\mathcal{G}}

\newcommand{\op}{\mathcal{O}}

\newcommand{\vev}[1]{\left\langle #1 \right\rangle}

\newcommand{\bfk}{\mathbf{k}}
\newcommand{\bfx}{\mathbf{x}}

\newcommand{\comb}[2]{\begin{pmatrix} #1\\#2\end{pmatrix}} 

\newcommand{\fconn}[2]{\mathcal{G}^{#1}_{\;\,{#2}}}

\newcommand{\bp}{\bar{\phi}}

\newcommand{\fa}{\mathfrak{a}}
\newcommand{\fb}{\mathfrak{b}}
\newcommand{\fe}{\mathfrak{e}}

\newcommand{\adom}{\hat{\fa}}
\newcommand{\asub}{\check{\fa}}

\begin{document}
\begin{flushright}
NA-DSF-27/2008
\end{flushright}
\vspace{1ex}

\begin{center}
\textbf{\Large Spectral Functions in Holographic Renormalization Group Flows}\\[4ex]%
Wolfgang M\"uck\\[3ex]%
\textit{Dipartimento di Scienze Fisiche, Universit\`a degli Studi di
  Napoli ``Federico II''\\ and INFN, Sezione di Napoli --- via Cintia, 80126 Napoli, Italy}\\%
E-mail: \texttt{mueck@na.infn.it}
\end{center}
\vspace{1ex}
\begin{abstract}
The spectrum of two-point functions in a holographic renormalization group flow from an ultraviolet (UV) to an infrared (IR) conformal fixed point is necessarily continuous. For a toy model, the spectral function does not only show the expected UV and IR behaviours, but other interesting features such as sharp peaks and oscillations in the UV. The spectral functions for the $SU(3) \times U(1)$ flow in $AdS_4/CFT_3$ and the $SU(2) \times U(1)$ flow in $AdS_5/CFT_4$ are calculated numerically. They exhibit a simple cross-over behaviour and reproduce the conformal dimensions of the dual operators in the UV and IR conformal phases.
\end{abstract}

\vspace{3ex}

\section{Introduction}
\label{intro}

In the AdS/CFT correspondence, holographic renormalization group (RG) flows provide the dual gravitational description of strongly coupled gauge theories, the conformal invariance of which is broken either by an explicit operator deformation or by some non-zero operator vacuum expectation value. The calculation of correlation functions in such field theories via holographic renormalization \cite{Bianchi:2001kw,Martelli:2002sp,Papadimitriou:2004rz} is an elegant and powerful method to obtain physically interesting quantities such as the mass spectra of glueballs \cite{Bianchi:2001de,Mueck:2001cy} and mesons (see \cite{Erdmenger:2007cm} for a review), scattering amplitudes \cite{Bianchi:2003ug,Mueck:2004qg}, as well as hydrodynamic transport coefficients of the high-temperature plasma phase \cite{Policastro:2001yc,Policastro:2002se,Policastro:2002tn}. Holographic RG flows have been used as gravity duals of high-$T_c$ superconductors \cite{Hartnoll:2008vx,Hartnoll:2008hs,Gubser:2008wv,Gubser:2008wz} and for constructing bottom-up AdS/QCD models \cite{Gursoy:2007cb,Gursoy:2007er,Zeng:2008sx}.
Recently, progress has also been made in non-AdS/non-CFT cases, \eg in the Klebanov-Strassler background \cite{Berg:2005pd,Berg:2006xy,Benna:2007mb}.

The present paper deals with holographic RG flows that interpolate between an ultraviolet (UV) and an infrared (IR) fixed point. In particular, we consider the gravity dual of the $SU(2)\times U(1)$ Leigh-Strassler flow \cite{Freedman:1999gp,Pilch:2000fu} and its three-dimensional cousin \cite{Ahn:2000aq,Ahn:2000mf,Corrado:2001nv}, which has been identified recently as a $SU(3)\times U(1)$ mass deformation of the Bagger-Lambert theory \cite{Ahn:2008ya,Benna:2008zy,Ahn:2008gda,Klebanov:2008vq}. In both cases, the holographic RG flow involves two (active) scalar fields and is known only numerically. To our knowledge, correlation functions in these backgrounds have not been studied so far. We intend to approach such calculations in the present paper by considering the two-point functions of the operators that are dual to the active scalars. In particular, we shall calculate the (eigenvalues of) the spectral function (matrix) and confirm that the IR physics of the conformal field theory is determined by the conformal operators living at the IR fixed point.

The following simple argument shows that the spectrum of two-point functions in any RG flow to an IR fixed point with a gravity dual must be continuous. In AdS/CFT, the bulk dual of a mass state is obtained by imposing two conditions on the solution of the linearized bulk equations of motion, namely  regularity in the bulk interior and absence of the dominant asymptotic mode (implying integrability). Typically, if there are regular and singular bulk solutions, these two conditions can be satisfied only for discrete mass values. If the deep bulk interior is described by an AdS geometry corresponding to the IR fixed point, the two deep bulk behaviours of a scalar field will be given by
\begin{equation}
   \label{intro:phi.IR}
  z\to\infty: \quad \phi_{\text{reg}} \sim z^{(d-1)/2} \e{-kL z}~, \quad 
    \phi_{\text{sing}} \sim z^{(d-1)/2} \e{kL z}~,
\end{equation}
where $d$ is the boundary dimension, $L$ is the IR AdS radius, and $k=(k^2)^{1/2}$. For generic complex momentum $k^2$ we choose the square root such that $\re k>0$, so that the modes are regular or singular as indicated by the subscripts. However, for $k^2=-m^2$ one has $\re k=0$, and both modes are wildly oscillating for large $z$. Hence, the regularity condition cannot be imposed, and the mass spectrum will be continuous with non-normalizable pseudo-states, as usual.

Let us now outline the rest of the paper and summarize the results. In Sec.~\ref{toy}, we start by considering a toy model for a RG flow between an UV and an IR fixed point consisting of two patches of AdS space with different radii glued together. We choose the masses of the scalar fields such that the bulk solutions are simple. Moreover, the UV conformal dimension of the dual operator coincides with those in the AdS$_4$/CFT$_3$ flow we study later. We verify that the spectral function exhibits the scaling behaviours in the UV (large mass) and the IR (small mass) that one expects for the operators living at the fixed points. In addition, we find interesting features in the spectral function such as sharp peaks and UV oscillations, which are are related to the discrete mass spectrum in the hard-wall model. Such features have been observed and interpreted as quasinormal modes in thermal spectral functions \cite{Erdmenger:2007ja}.

In Sec.~\ref{specRG}, we present the technical tools that we need for calculating spectral functions in holographic RG flows. We start by reviewing the spectral function in many-particle systems, continue with a review of the equations governing the bulk dynamics in holographic RG flow backgrounds and then present the numerical strategy for the calculations. In Secs.~\ref{a4c3} and \ref{a5c4}, the $SU(3)\times U(1)$ flow in AdS$_4$/CFT$_3$ and the $SU(2)\times U(1)$ flow in AdS$_5$/CFT$_4$ are considered, respectively. In both cases, we review the RG flow backgrounds and present our numerical calculations of the eigenvalues of the density function. Our results indicate that the deformed CFTs exhibit a simple cross-over behaviour from the UV to the IR regime. In the UV regime, the CFTs behave as the undeformed ones, while the IR behaviour is dominated by the operators that live at the IR fixed point.

\section{Toy Model}
\label{toy}

Let us consider the following simple toy model consisting of two patches of $(d+1)$-dimensional AdS bulk space-time glued together at some fixed radial coordinate. We use an AdS metric of the form
\begin{equation}
 \label{toy:metric}
  \rmd s^2 = \rmd r^2 + \e{2A(r)} \eta_{ij} \rmd x^i \rmd x^j~,
\end{equation}
where
\begin{equation}
 \label{toy:warp}
  A(r) = r/L +A_0~,
\end{equation}
$L$ being the AdS length scale, and $A_0$ an arbitrary constant. The equation of motion for a free massive scalar field can written in the form
\begin{equation}
\label{toy:eom}
  \left[ \left( \partial_r + \frac{d+2\lambda}{2L} \right) 
     \left( \partial_r + \frac{d-2\lambda}{2L} \right) -k^2 \e{-2A(r)} \right] \phi =0~.
\end{equation}
The scalar's mass $m$ is determined by the well known relation $\lambda= \sqrt{d^2/4+m^2L^2}$, while the dimension of the dual operator is, in the case of regular boundary conditions, $\Delta=d/2+\lambda$.
After introducing the dimensionless variables
\begin{equation}
\label{toy:rescale}
   \tik^2= k^2L^2\e{-2A_0}~,\qquad z= \e{-r/L}~,
\end{equation}
\eqref{toy:eom} takes a standard form and has the solutions
\begin{equation}
\label{toy:eom.sols}
  \phi_{\text{reg}} = z^{d/2} \BesK_\lambda(\tik z)~, \qquad 
  \phi_{\text{sing}} = z^{d/2} \BesI_\lambda(\tik z)~,
\end{equation}
where $\BesK_\lambda$ and $\BesI_\lambda$ denote modified Bessel functions. 
Here and henceforth, $\tik$ is defined as the square root of $\tik^2$ with $-\pi/2 < \arg \tik \leq \pi/2$.

In the following, let us consider a UV region, $r>0$, with parameters $L_\UV =1$ and $A_{0,\UV}=0$, in which lives a scalar field with $\lambda_\UV=1/2$. This choice coincides with the UV behaviour of the 3d RG flow that we study in Sec.~\ref{a4c3}. For $\lambda=1/2$, the modified Bessel functions in \eqref{toy:eom.sols} are elementary. Hence, the general solution of \eqref{toy:eom} becomes
\begin{equation}
  \label{toy:phi.UV}
  \pUV = z^{(d-1)/2} \left[ \sour \cosh(kz) +\frac{\resp}{k} \sinh(kz) \right]~,
\end{equation}
where $\sour$ and $\resp$ are called the source and response coeffiencts, respectively. They determine the connected two-point function of the dual operator by
\begin{equation}
   \label{toy:prop.def}
   \prop (\bfk) = \frac{\resp(\bfk)}{\sour(\bfk)}~,
\end{equation}
in terms of which the spectral function $\rho$ is defined as
\begin{equation}
   \label{toy:spec.fun}
   \rho = -2 \im \prop~,\quad \text{with } k^2=-m^2+i0^+~.
\end{equation}

In the IR region, $r<0$, we set $L_\IR=l$, while the continuity of the metric at $r=0$ imposes $A_{0,\IR}=0$.
Let us choose $\lambda_\IR=1/2+n$ for some positive integer $n$, so that the solutions \eqref{toy:eom.sols} are again elementary.
Imposing the regularity condition for $r\to-\infty$ (the deep interior) selects the solution
\begin{equation}
  \label{toy:phi.IR}
  \pIR = z^{d/2} \BesK_{n+1/2}(kl z) 
  \sim z^{(d-1)/2} \e{-kl z} \sum\limits_{j=0}^n \frac{(n+j)!}{j!(n-j)! (2kl z)^j}~.
\end{equation}
The overall numerical factor is irrelevant for what follows, as it drops out of \eqref{toy:prop.def}.

The source and response coefficients, $\sour$ and $\resp$, are determined by appropriate matching conditions at $r=0$, \ie $z=1$. From \eqref{toy:eom} and \eqref{toy:rescale} follows that the matching conditions are
\begin{equation}
 \label{toy:match}
  z=1:\qquad \pUV=\pIR~,\qquad 
  -\partial_z\pUV +\frac{d-1}2 \pUV = 
   \frac1l \left[ -\partial_z \pIR + \left( \frac{d-1}2 -n \right) \pIR \right]~.
\end{equation}
They give rise to
\begin{align}
  \label{toy:match.1}
  \frac12 \left( \sour +\frac{\resp}k \right) \e{k} &=
     \e{-kl} \sum\limits_{j=1}^n \frac{(n+j-1)!}{(j-1)!(n-j)! (2kl)^j}~, \\
  \label{toy:match.2}
  \frac12 \left( \sour -\frac{\resp}k \right) \e{-k} &=
     \e{-kl} \left[1+ n\sum\limits_{j=1}^n \frac{(n+j-1)!}{j!(n-j)! (2kl)^j} \right]~,
\end{align}
so that one obtains, using \eqref{toy:prop.def},
\begin{equation}
  \label{toy:prop.implicit}
  \frac{k+\prop}{k-\prop} = \e{-2k} 
   \frac{\sum\limits_{j=1}^n \frac{(n+j-1)!}{(j-1)!(n-j)! (2kl)^j} }{1+
      n\sum\limits_{j=1}^n \frac{(n+j-1)!}{j!(n-j)! (2kl)^j}}~.
\end{equation}
Curiously, the boundary dimension $d$ has dropped out. Defining $\alpha$ by
\begin{equation}
 \label{toy:redef.1}
  \frac{k+\prop}{k-\prop} = \e{-2\alpha}\quad \Rightarrow \quad \frac{\prop}{k} = -\tanh \alpha~,
\end{equation}
and, similarly, $y$ and $\beta$ by 
\begin{equation}
 \label{toy:redef.2}
  \frac{\sum\limits_{j=1}^n \frac{(n+j-1)!}{(j-1)!(n-j)! (2kl)^j} }{1+
      n\sum\limits_{j=1}^n \frac{(n+j-1)!}{j!(n-j)! (2kl)^j}} = \frac{1-y}{1+y} = \e{-2\beta}
  \quad \Rightarrow \quad y= \tanh \beta~,
\end{equation}
\eqref{toy:prop.implicit} yields
\begin{equation}
 \label{toy:prop.explicit}
  \prop(\bfk) = -k \frac{\tanh k + y}{1+y\tanh k}~,
\end{equation}
where $y$ is determined from \eqref{toy:redef.2} as 
\begin{equation}
 \label{toy:y.1}
  y = \left[\sum\limits_{j=0}^n \frac{(n-1+j)!}{(n-1-j)!j! (2kl)^j}\right]
    \left[\sum\limits_{j=0}^n \frac{(n+j)!}{(n-j)!j! (2kl)^j} \right]^{-1}
    = \frac{\BesK_{n-1/2}(kl)}{\BesK_{n+1/2}(kl)}~.
\end{equation}

In the following, let us extract the UV and IR behaviours of the spectral function and verify that they match with the spectral functions for operators of dimensions $\Delta_\UV=(d+1)/2$ and $\Delta_\IR = (d+1)/2+n$, respectively. In the UV, \ie for large $|k|$, $y$ is approximately $y \approx 1 -n/(kl)$,
such that \eqref{toy:prop.explicit} becomes
\begin{equation}
\label{toy:prop.UV}
   \UV: \quad \prop(\bfk) \approx -k +\frac{n}{l} \e{-2k}~.
\end{equation}
For $\re k>0$, only the first term on the right hand side remains in the UV limit, which is precisely the propagator of a $\Delta=2$ operator, as expected. In addition, the second term gives rise to oscillations in the spectral function \eqref{toy:spec.fun}, which becomes
\begin{equation}
\label{toy:spec.UV}
   \UV: \quad \rho(m^2) \approx 2\sqrt{m^2} +\frac{2n}{l} \sin(2\sqrt{m^2})~.
\end{equation}

To obtain the IR (small $m$) behaviour of the spectral function, we start by inserting \eqref{toy:prop.explicit} into \eqref{toy:spec.fun},
\begin{equation}
  \label{toy:spec.dens}
  \rho(m^2) = -ik(y+y^\ast) \frac{1+|\tanh k|^2}{|1+y\tanh k|^2}~, \qquad (k=im)~.
\end{equation}
Since, on the imaginary axis, $k^\ast=-k$, only those terms of $y$ that are even in $k$ contribute to $(y+y^\ast)$. To find these, let us rewrite \eqref{toy:y.1} as
\begin{equation}
\label{toy:y.2}
  y = \frac{kl}{2n-1} \frac{P_{n-1}(kl)}{P_n(kl)}~,
\end{equation}
where the polynomials $P_n(x)$ are defined by
\begin{equation}
 \label{toy:poly}
  P_n(x)=\sum\limits_{j=0}^n \frac{\comb{n}{j}}{\comb{2n}{j}} \frac{(2x)^j}{j!}~.
\end{equation}
It is straightforward to show that the $P_n$ satisfy the recursion relation
\begin{equation}
  \label{toy:recursion}
  P_n(x) = P_{n-1}(x) + \frac{x^2}{(2n-1)(2n-3)} P_{n-2}(x) \qquad (n\geq2)~,
\end{equation}
while the cases $n=0$ and $n=1$ are simply $P_0(x)=1$ and $P_1(x)=1+x$, respectively. Hence,
\begin{equation}
  \label{toy:p.frac}
   \frac{P_{n-1}(x)}{P_n(x)} = \frac1{1+\frac{x^2}{(2n-1)(2n-3)}\frac{P_{n-2}(x)}{P_{n-1}(x)}}~,
\end{equation}
which can be used recursively until one hits the term $P_0(x)/P_1(x)$. Therefore, expanding \eqref{toy:p.frac} for small $x$, the first odd term one encounters is the one that stems from $P_0(x)/P_1(x)\approx 1-x$. Thus, one has
\begin{equation}
  \label{toy:p.frac.expand}
   \frac{P_{n-1}(x)}{P_n(x)} = \text{even terms} 
   + \frac{(-1)^n (2n-1) x^{2n-1}}{[(2n-1)!!]^2} +\cdots~,
\end{equation}
so that \eqref{toy:y.2} leads to\footnote{This result can also be obtained, by similar reasoning, considering $y$ in terms of the modified Bessel functions [see \eqref{toy:y.1}] and using the recursion relation $\BesK_{\nu-1}(x)-\BesK_{\nu+1}(x)=-2\nu/x \BesK_\nu(x)$.}
\begin{equation}
  \label{toy:re.y.expand}
  y+y^\ast = \frac{2 (-1)^n (kl)^{2n}}{[(2n-1)!!]^2} + \cdots~.
\end{equation}
Finally, inserting \eqref{toy:re.y.expand} into \eqref{toy:spec.dens} yields
\begin{equation}
  \label{toy:spec.IR}
  \IR:\quad \rho(m^2) = \frac{2l^{2n}(m^2)^{n+1/2}}{[(2n-1)!!]^2} +\cdots~.
\end{equation}
The power behaviour $m^{2n+1}$ is expected for a dimension $\Delta_\IR= (d+1)/2 +n$ operator.

\begin{figure}[th]
(a)\includegraphics[width=0.47\textwidth]{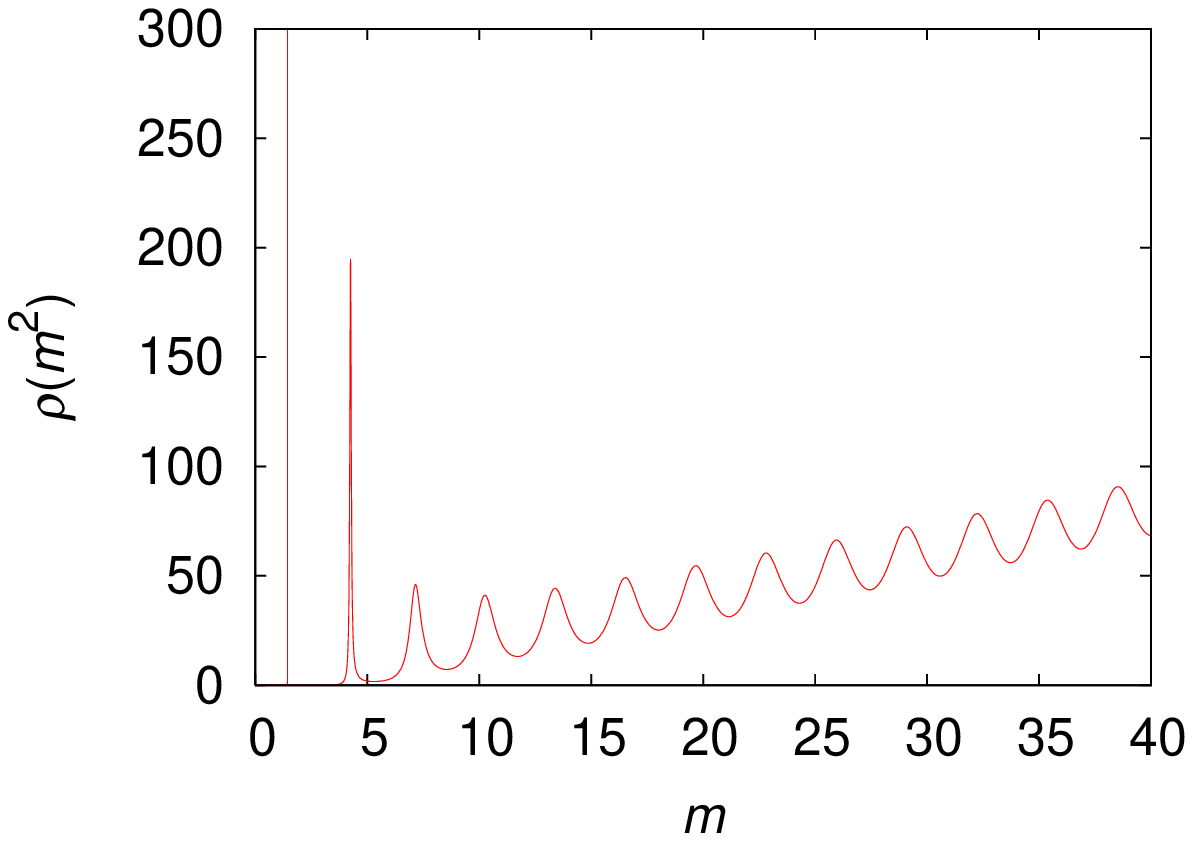} \hfill
(b)\includegraphics[width=0.47\textwidth]{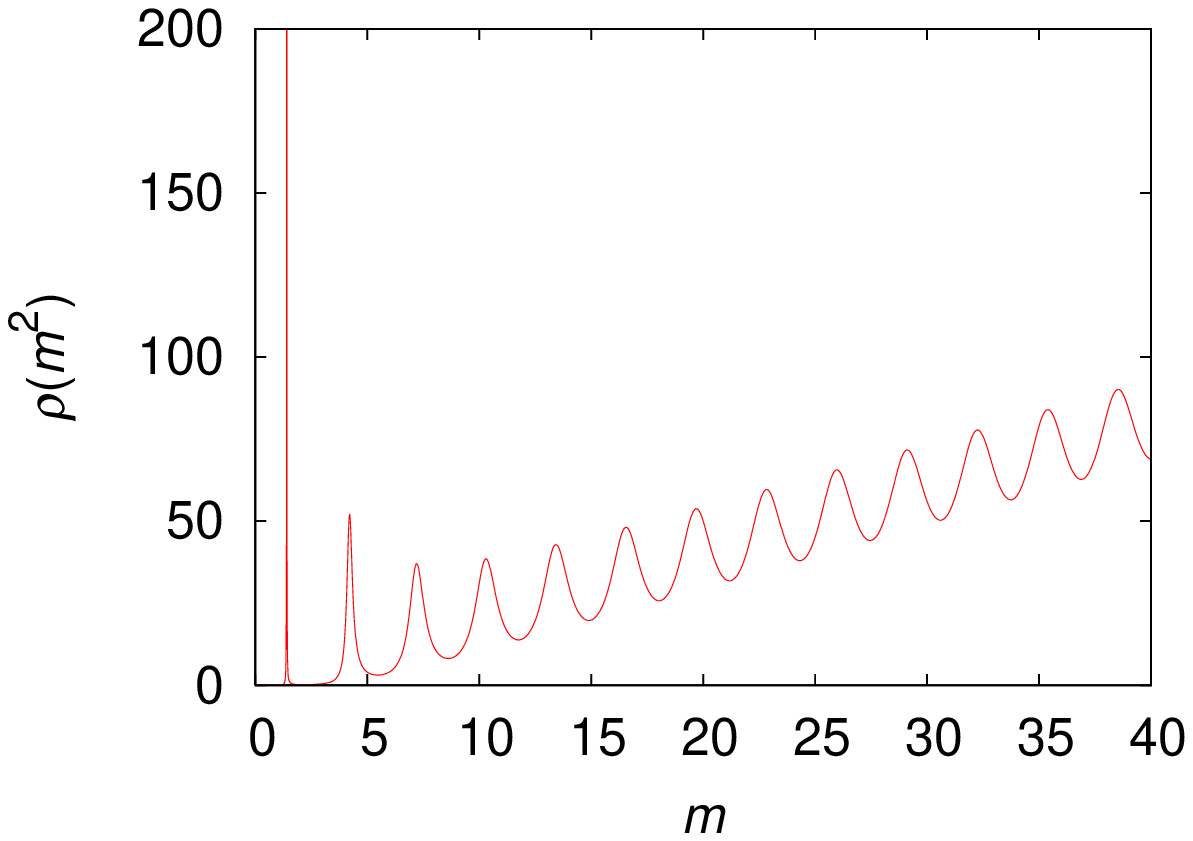}
\caption{Spectral function $\rho(m^2)$ for parameters (a) $n=5$, $l=4/5$ and (b) $n=2$, $l=1/3$. The sharp peaks are due to poles of the right hand side of \eqref{toy:prop.explicit} that lie in the unphysical region, but close to the imaginary axis $k=im$. The first peak of every plot is actually much taller, but the $y$-range has been cut off to show the other features. The UV behaviour \eqref{toy:spec.UV} with oscillations is clearly visible. \label{toy:spec.fig}}
\end{figure}

Another interesting limit is $l\to0$, which describes the set-up of a hard-wall cut-off. In this limit, \eqref{toy:y.2} and \eqref{toy:poly} imply $y=0$, so that \eqref{toy:prop.explicit} becomes
\begin{equation}
 \label{toy:l0.prop}
  l=0:\quad \prop(\bfk) =-k \tanh k~.
\end{equation}
In this case, the non-vanishing spectral function stems from the poles of $\tanh k$, which lie just on the imaginary axis. Setting $k=im+\varepsilon$, one easily finds the discrete spectrum
\begin{equation}
 \label{toy:l0.dens}
  l=0:\quad \rho(m^2) = 2 \pi m \sum\limits_{j=0}^\infty \delta \left(m-\frac{2j+1}{2}\pi \right)~.
\end{equation}
In the general case, the right hand side of \eqref{toy:prop.explicit} has an infinite number of poles on the left side of the $k$-plane, \ie in the unphysical region $\im k<0$.\footnote{Poles in the physical region $\re k >0$ would give rise to poles of $\prop(k^2)$, which in turn would be a signal of unstable states, or resonances. The absence of such poles implies the absence of resonances.}
The precise locations of the poles depend on $n$ and $l$, but the limit $l\to 0$ moves them onto the imaginary axis. For non-zero $l$, though, a number of poles tend to lie very close to the imaginary axis leading to a sharply peaked spectral function. Moreover, the UV oscillations in the spectral function \eqref{toy:spec.UV} stem from the poles in the $k$-plane. It is an interesting fact that these features have also been found in the context of thermal spectral functions at finite baryon density \cite{Erdmenger:2007ja}. The results are illustrated in Fig.~\ref{toy:spec.fig}.

\section{Spectral Functions in Holographic RG flows}
\label{specRG}

\subsection{The Spectral Function}
\label{specdens}

The spectral representation of two-point functions is, of course, well known, but readers might remember only the simple case of a single operator. As the general, many-particle case will be needed later, we briefly review it in this section.

Consider the two-point correlation function between operators $\op_i$ and $\op_j$,
\begin{equation}
 \label{specdens:green}
  \prop_{ij}(\bfk) = \int \rmd^d x \vev{\op_i(\bfx) \op_j^\dagger(0)} \e{i\bfk \cdot \bfx}~,
\end{equation}
and its spectral representation
\begin{equation}
 \label{specdens:spec.rep}
 \prop_{ij}(\bfk) = \int\limits_0^\infty \frac{\rmd m^2}{2\pi} \frac{\rho_{ij}(m^2)}{k^2+m^2}~,
\end{equation}
where $\rho_{ij}$ is the spectral function (matrix), often also called spectral density. For brevity, let us suppress the operator indices in what follows and regard $\prop(\bfk)$ and $\rho(m^2)$ as matrices.
Considering complex $k^2$, the two-point function may have poles and branch cuts along the negative real axis corresponding, respectively, to the discrete and continuous elements of the spectrum. The spectral function $\rho$ is determined by the discontinuity of the correlation function across the branch cut,
\begin{equation}
 \label{specdens:spec.dens}
  \rho(m^2) = i \left( \prop_+ -\prop_- \right)~,
\end{equation}
where the subscripts $+$ and $-$ indicate the sign of the (infinitesimal) imaginary part of $k^2=-m^2+ i0^\pm$. In addition, the correlation function \eqref{specdens:green} satisfies the analyticity property 
\begin{equation}
 \label{specdens:anal.prop}
  \left[\prop(k^2)\right]^\dagger = \prop[(k^2)^\ast]~,
\end{equation}
which implies
\begin{equation}
 \label{specdens:anal.pm}
  \prop_- = \prop_+^\dagger~.
\end{equation}
Combining \eqref{specdens:anal.pm} with \eqref{specdens:spec.dens} yields
\begin{equation}
 \label{specdens:rho.pm}
 \rho(m^2) = i \left( \prop_+ -\prop_+^\dagger \right)~,
\end{equation}
which shows that $\rho$ is a hermitian matrix with real eigenvalues. Eq.~\eqref{specdens:rho.pm} reduces to the well known $\rho= -2 \im \prop_+$ for the case of a single operator.

\subsection{Bulk Dynamics}
\label{bdyn}

Let us start by reviewing the equations governing the dynamics of the bulk fields \cite{Mueck:2004ih,Berg:2005pd}, which encode the information about two-point functions in holographic renormalization group flows. Then, we will introduce a change of variables that further enhances the system eliminating the need to calculate the warp factor.

The systems we consider are of fake SUGRA type with actions of the form
\begin{equation}
 \label{bdyn:action}
  S= \int \rmd^{d+1}x \sqrt{g} \left[ -\frac14 R +\frac12 G_{ab} \partial_\mu \phi^a \partial^\mu \phi^b +V(\phi)\right] +S_b~,
\end{equation}
where the potential $V(\phi)$ is given in terms of a superpotential $W(\phi)$ by
\begin{equation}
 \label{bdyn:potential}
  V(\phi) = \frac12 G^{ab} W_a W_b -\frac{d}{d-1} W^2~.
\end{equation}
We will not specify the boundary terms $S_b$ in \eqref{bdyn:action}, as they do not affect the bulk dynamics, although they are important for holographic renormalization. Our notation agrees with \cite{Berg:2005pd}. In particular, field indices are covariantly lowered and raised with the sigma-model metric $G_{ab}$ and its inverse, $G^{ab}$, respectively; $W_a=\partial_a W=\partial W(\phi)/\partial \phi^a$; and covariant derivatives with respect to the fields are indicated by $D_a$ or by a ``$|$'' preceding the index, as in $W_{a|b} = D_b W_a = \partial_b W_a - \fconn{c}{ab} W_c$, $\fconn{c}{ba}$ being the Christoffel symbol for the metric $G_{ab}$.

Holographic renormalization group flows are described by domain wall backgrounds of the form
\begin{equation}
 \label{bdyn:bg}
  \rmd s^2 = \rmd r^2 +\e{2A(r)} \eta_{ij}\rmd x^i \rmd x^j~, \qquad \phi^a= \bp^a(r)~,
\end{equation}
which satisfy the BPS equations
\begin{equation}
 \label{bdyn:BPS}
  \partial_r A= -\frac2{d-1} W(\bp)~, \qquad \partial_r \bp^a = W^a(\bp)~.
\end{equation}

Linearized fluctuations around the domain wall background are best described in a gauge invariant fashion, in which the relevant, independent fields are the scalar fluctuations $\fa^a$ and the traceless transversal metric fluctuations, $\fe^i_j$. They satisfy the (linearized) equations of motion
\begin{equation}
 \label{bdyn:eom.a}
  \left[\left( D_r +\tilde{M}-\frac{2d}{d-1}W \right)\left( D_r -\tilde{M}\right) 
	+\e{-2A}\Box \right] \fa =0
\end{equation}
and
\begin{equation}
 \label{bdyn:eom.e}
  \left[\left( \partial_r -\frac{2d}{d-1}W \right) \partial_r  +\e{-2A}\Box \right] \fe^i_j =0~,
\end{equation}
respectively. In \eqref{bdyn:eom.a}, we have omitted the field indices, $\tilde{M}$ denotes the matrix
\begin{equation}
 \label{bdyn:M.def}
  \tilde{M}^a{}_{b} = W^a{}_{|b} -\frac{W^a W_b}{W}~,
\end{equation}
and $D_r$ is the background covariant derivative
\begin{equation}
 \label{bdyn:Dr}
  D_r \fa^a = \partial_r \fa^a + \fconn{a}{bc} W^b \fa^c~.
\end{equation}
For more details, we refer the reader to the original papers \cite{Bianchi:2003ug,Mueck:2004ih,Berg:2005pd}.

Let us now introduce a change of variables, which facilitates the treatment of the bulk dynamics by eliminating the warp function $A(r)$. Introducing a new radial variable $\sigma$ by 
\begin{equation}
   \label{bdyn.rho.def}
	\sigma= A(r)~,
\end{equation}
where we implicitly assume that $A(r)$ is a monotonous function, or equivalently, that $W$ does not change sign along the RG flow, the BPS equations \eqref{bdyn:BPS} yield simply
\begin{equation}
   \label{bdyn:BPS.2}
	\partial_\sigma \bp^a = -\frac{d-1}2 \frac{W^a(\bp)}{W(\bp)}~,
\end{equation}
and the field equations \eqref{bdyn:eom.a} and \eqref{bdyn:eom.e} become
\begin{equation}
 \label{bdyn:eom.a2}
  \left[\left( D_\sigma +M+d \right)\left( D_\sigma -M\right) 
	+\e{-2\sigma}\frac{(d-1)^2}{4W^2} \Box \right] \fa =0
\end{equation}
and
\begin{equation}
 \label{bdyn:eom.e2}
  \left[\left( \partial_\sigma +d \right) \partial_\sigma  
	+\e{-2\sigma}\frac{(d-1)^2}{4W^2} \Box \right] \fe^i_j =0~,
\end{equation}
respectively. In \eqref{bdyn:eom.a2}, we have
\begin{equation}
 \label{bdyn:M2}
  M^a{}_{b} = -\frac{d-1}{2W} \tilde{M}^a{}_b = 
   -\frac{d-1}{2} \left( \frac{W^a{}_{|b}}{W} -\frac{W^a W_b}{W^2} \right)
  = -\frac{d-1}{2} D_b \left( \frac{W^a}{W} \right)~,
\end{equation}
and 
\begin{equation}
 \label{bdyn:Dsigma}
  D_\sigma \fa^a = -\frac{d-1}{2W} D_r \fa^a = \partial_\sigma \fa^a 
    + \fconn{a}{bc} (\partial_\sigma \bp^b) \fa^c~.
\end{equation}

As an aside, we observe that $\partial_\sigma \bp^a$ is a solution to the linearized field equation \eqref{bdyn:eom.a2}.

\subsection{Near Boundary Asymptotics and Pure AdS Case}
\label{asymp}

To extract the CFT data from the bulk dynamics, we make use of holographic renormalization \cite{Bianchi:2001kw,Martelli:2002sp,Papadimitriou:2004rz}. The near boundary behaviour (large $\sigma$) of the solutions of \eqref{bdyn:eom.a2} is determined by the UV fixed point, at which the matrix $M$ is chosen diagonal with eigenvalues $\lambda_i-d/2$. This follows from the expansion of $W$ around the fixed point, which for a single scalar $\phi$ reads \cite{Martelli:2002sp}
\begin{equation}
\label{asymp:W.exp}
	W(\phi) = -\frac{d-1}{2L} -\frac1{4L} \left(d-2\lambda \right)\phi^2 +\cdots~.
\end{equation}
The (UV) conformal dimensions of the dual operators $\op_i$ are related to the eigenvalues of $M$ by $\Delta_i=d/2+|\lambda_i|$. Notice that the $\lambda_i$ can be negative. 

A generic solution of \eqref{bdyn:eom.a2} can be decomposed as
\begin{equation}
\label{asymp:scal.decomp}
  \fa^a(\sigma)= \sour_i \, \adom^a_i(\sigma) + \resp_i\, \asub^a_i(\sigma)~,
\end{equation}
where $\adom_i$ and $\asub_i$ denote the dominant and sub-dominant series solutions, which are the duals of the CFT operators $\op_i$ of dimensions $\Delta_i$, and $\sour_i$ and $\resp_i$ are the source and response coefficients, respectively.\footnote{Note the change of notation with respect to \cite{Martelli:2002sp}, where the source and response coefficients were denoted by $\hat{}$ and $\check{}$, respectively.}
The leading terms of $\adom_i$ and $\asub_i$ agree with the respective counterparts in pure AdS,\footnote{In the special case $\Delta_i=d/2$, one has $\adom=\rho \e{-d\rho/2}+\cdots$.}
\begin{equation}
\label{asymp:dom.sub}
  \adom^a_i(\sigma) = \frac1{2\Delta_i-d} \e{-(d-\Delta_i)\sigma} \delta_i^a +\cdots~,\qquad
  \asub^a_i(\sigma) = \e{-\Delta_i \sigma} \delta^a_i +\cdots~.
\end{equation}
The somewhat unconventional normalization of the dominant solutions eliminates the proportionality factor $2|\lambda_i|$ in the standard relation between the exact one-point function and the response coefficients \cite{Klebanov:1999tb,Mueck:1999kk}. The sub-leading terms in \eqref{asymp:dom.sub} depend on the background and must be considered case by case.

To calculate the Green's function, one imposes a regularity condition on $\fa$. If $\fa$ has $n_s$ components, then there are $n_s$ independent regular solutions. Hence, $\sour$ and $\resp$ can be considered as $n_s \times n_s$ matrices, the first index labelling the regular solutions and the second index the asymptotic (sub-)dominant ones. Then, the Green's function (matrix) has the elegant form
\begin{equation}
\label{asymp:green}
  \prop = \sour^{-1}\cdot \resp~,
\end{equation}
where the matrix multiplication sums over the index labelling the regular solutions.

For a scalar field in pure AdS with generic mass ($\lambda$ non-integer), one has the standard result \cite{Gubser:1998bc}
\begin{equation}
\label{asymp:green.ads}
  \prop(\bfk) = \frac{\Gamma(-|\lambda|)}{2\Gamma(1+|\lambda|)} 
   \left( \frac{k}2 \right)^{2|\lambda|}~,
\end{equation}
and the spectral function \eqref{specdens:rho.pm} is\footnote{In contrast to \eqref{asymp:green.ads}, formula \eqref{asymp:rho} is valid also for integer $\lambda$.}
\begin{equation}
\label{asymp:rho}
  \rho(m) = \frac{\pi}{[\Gamma(1+|\lambda|)]^2} \left( \frac{m}2 \right)^{2|\lambda|}~.
\end{equation}

\subsection{Numerical Strategy}
\label{num:strat}
The numerical strategy we employ for calculating two-point correlation functions and their spectral functions in holographic renormalization group flows between two conformal fixed points closely follows what one would do for an analytic calculation. The starting point is the equation of motion \eqref{bdyn:eom.a2}, which we rewrite as a system of first-order ordinary differential equations (ODEs),
\begin{equation}
\label{num:eom}
  D_\sigma \begin{pmatrix}
         \fa \\ \fb
      \end{pmatrix} =
  \begin{pmatrix}
     M & \idmat \\ \left[\frac{d-1}{2W} \e{-\sigma} \right]^2 k^2 & -M-d
  \end{pmatrix}
  \begin{pmatrix}
     \fa \\ \fb
  \end{pmatrix}~,
\end{equation}
where some auxiliary components $\fb$ have been introduced, and $\idmat$ is a unit matrix. In this paper, we deal with canonical scalar fields, so that the derivative is simply $D_\sigma=\partial_\sigma$.
As we must consider complex $k^2=-m^2 +i0^+$, with some tiny imaginary part,\footnote{As we mentioned in the introduction, an IR AdS region in the bulk interior implies that $\re k >0$ is a necessary condition for distinguishing between regular and singular solutions.} we must also treat the real and complex components of the scalars as independent, so that, for $n_s$ (complex) scalar fields, \eqref{num:eom} is a system consisting of $4n_s$ ODEs. Hence, decomposing each scalar field as
\begin{equation}
   \label{num:a.decomp}
	\fa \to \begin{pmatrix}
	           \re \fa \\ \im \fa
	        \end{pmatrix}~,
\end{equation}
the complex momentum $k$ becomes a $2\times 2$ matrix,
\begin{equation}
   \label{num:k.decomp}
	k \to \begin{pmatrix}
	           \re k & -\im k \\
		   \im k & \re k
	        \end{pmatrix}~.
\end{equation}

We are interested in the solutions of \eqref{num:eom} that are regular in the bulk interior, \ie which behave as $\phi_\text{reg}$ in \eqref{intro:phi.IR} to leading order. However, using the exponential damping to impose this behaviour is not practical, because $\re k$ is tiny, and the bulk solution typically undergoes wild fluctuations in the region dominated by the $z^{(d-1)/2}$ factor before reaching the physically interesting domain wall. Thus, it is better to define the fields
\begin{equation}
   \label{num:a.IR.def}
   \fa'= \e{kL_\IR(\e{-\sigma}-\e{-\sigma_0})} \fa~,\qquad 
   \fb'= \e{kL_\IR(\e{-\sigma}-\e{-\sigma_0})} \fb~,
\end{equation}
which satisfy the field equation
\begin{equation}
\label{num:eom.IR}
  D_\sigma \begin{pmatrix}
         \fa' \\ \fb'
       \end{pmatrix} =
  \begin{pmatrix}
     M -kL_\IR\e{-\sigma} & \idmat \\ 
     \left[\frac{d-1}{2W} \e{-\sigma}\right]^2 k^2 & -M-d -kL_\IR\e{-\sigma}
  \end{pmatrix}
  \begin{pmatrix}
     \fa' \\ \fb'
  \end{pmatrix}~.
\end{equation}
We numerically integrate \eqref{num:eom.IR} for $\sigma<\sigma_0$ and then continue with \eqref{num:eom} for $\sigma>\sigma_0$, with $\sigma_0$ chosen in vicinity of the domain wall. The initial condition that we impose in the IR can be determined from the leading and next-to leading behaviour of $\fa'$, which is
\begin{equation}
 \label{num:a.IR.beh}
	\sigma \to -\infty: \qquad \fa' = \e{-\frac{d-1}2 \sigma} \left( 1 + \gamma \e{\sigma} + \cdots \right) \fa'_0~,
\end{equation}
where $\fa'_0$ is a constant field vector, and the matrix $\gamma$ is determined from \eqref{num:eom.IR} as
\begin{equation}
 \label{num:gamma}
  \gamma = \frac{k^\ast}{2L_\IR |k|^2} \left[ \frac{d^2-1}{4} +D_\sigma M + (M+d) M \right]~.
\end{equation}
There are $2n_s$ independent initial conditions, which give rise to $2n_s$ independent regular solutions. The field redefinitions \eqref{num:a.IR.def} have been chosen such that the vector $(\fa'\, \fb')$ matches smoothly with $(\fa\, \fb)$ at $\sigma=\sigma_0$, but omitting the term containing $\sigma_0$ would just amount to a multiplication by an irrelevant (complex) normalization factor.

In the UV region, the source and response coefficients, $\sour$ and $\resp$, respectively, can be extracted from the numerical solutions by using the decomposition \eqref{asymp:scal.decomp} into dominant and sub-dominant terms. Then, the Green's function is obtained from \eqref{asymp:green} as a $2n_s \times 2n_s$ matrix consisting of $n_s \times n_s$ blocks of $2 \times 2$ matrices of the form
\[
   \begin{pmatrix}
    \re \prop_{ij} & \im \prop_{ij} \\
    -\im \prop_{ij} & \re \prop_{ij}
   \end{pmatrix}~.
\]
Finally, \eqref{specdens:rho.pm} yields the spectral function (matrix) $\rho$, the eigenvalues of which are then calculated.

\begin{figure}[th]
\begin{center}
\includegraphics[width=0.5\textwidth]{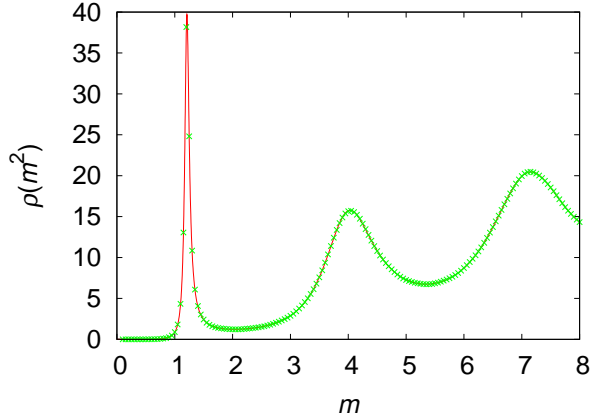}
\caption{Spectral density of the toy model with parameters $d=3$, $L_\IR=0.8$, $n=2$. The numerical solution (points) agrees with the exact solution (line).\label{num:fig}}
\end{center}
\end{figure}

To test the numerical strategy described above, we have numerically calculated the spectral density for the toy model studied in Sec.~\ref{toy}. In particular, we have considered the parameters $d=3$, $L_\IR=0.8$ and $n=2$, \ie the scalar in the IR region is dual to an operator of dimension $\Delta_\IR=4$. The integration range was chosen as $-4\leq r \leq 8$, and the IR and UV regions are glued together at $r=0$.

The exact solution for the spectral function is given by \eqref{toy:spec.fun} and  \eqref{toy:prop.explicit}. The results are depicted in Fig.~\ref{num:fig}, and the agreement between the exact and the numerical solutions is evident.

\section{$SU(3)\times U(1)$ RG flow in $d=3$}
\label{a4c3}
\subsection{Fixed Points and Background Solution}
\label{a4c3:bg.sec}

\begin{figure}[th]
(a)\includegraphics[width=0.47\textwidth]{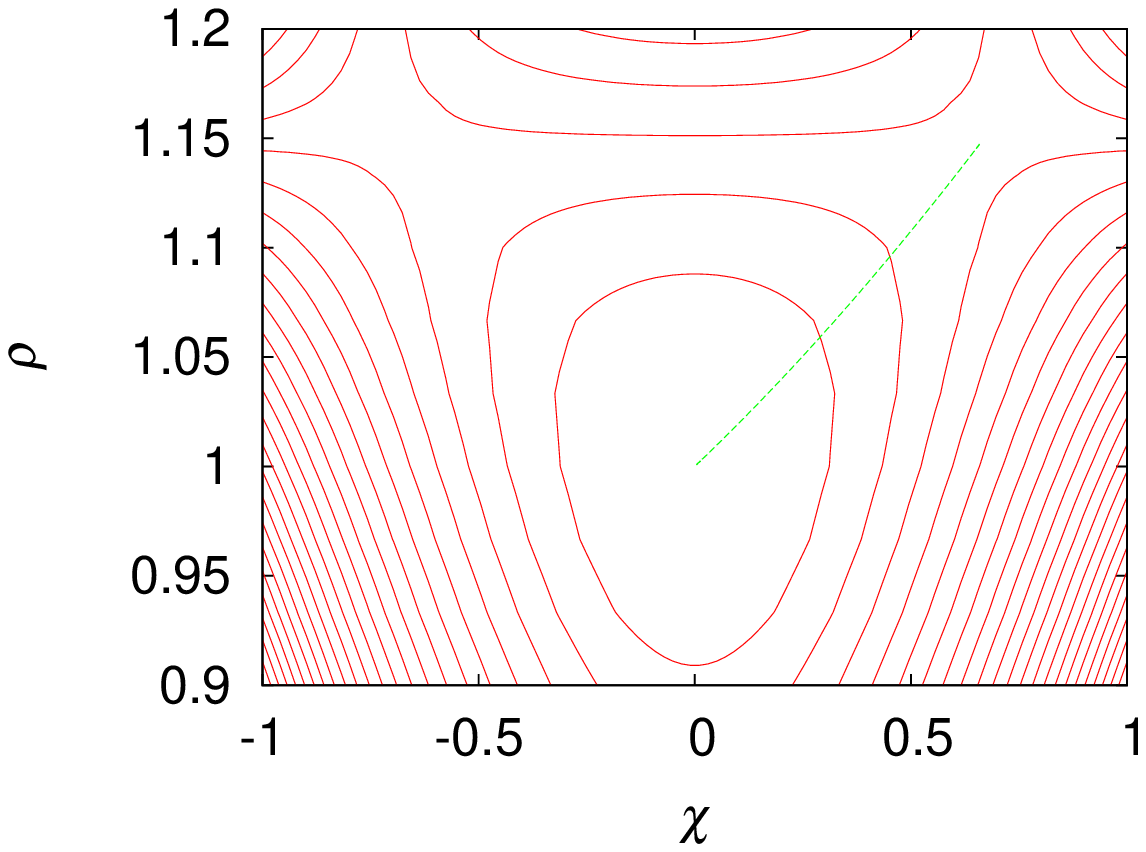} \hfill
(b)\includegraphics[width=0.47\textwidth]{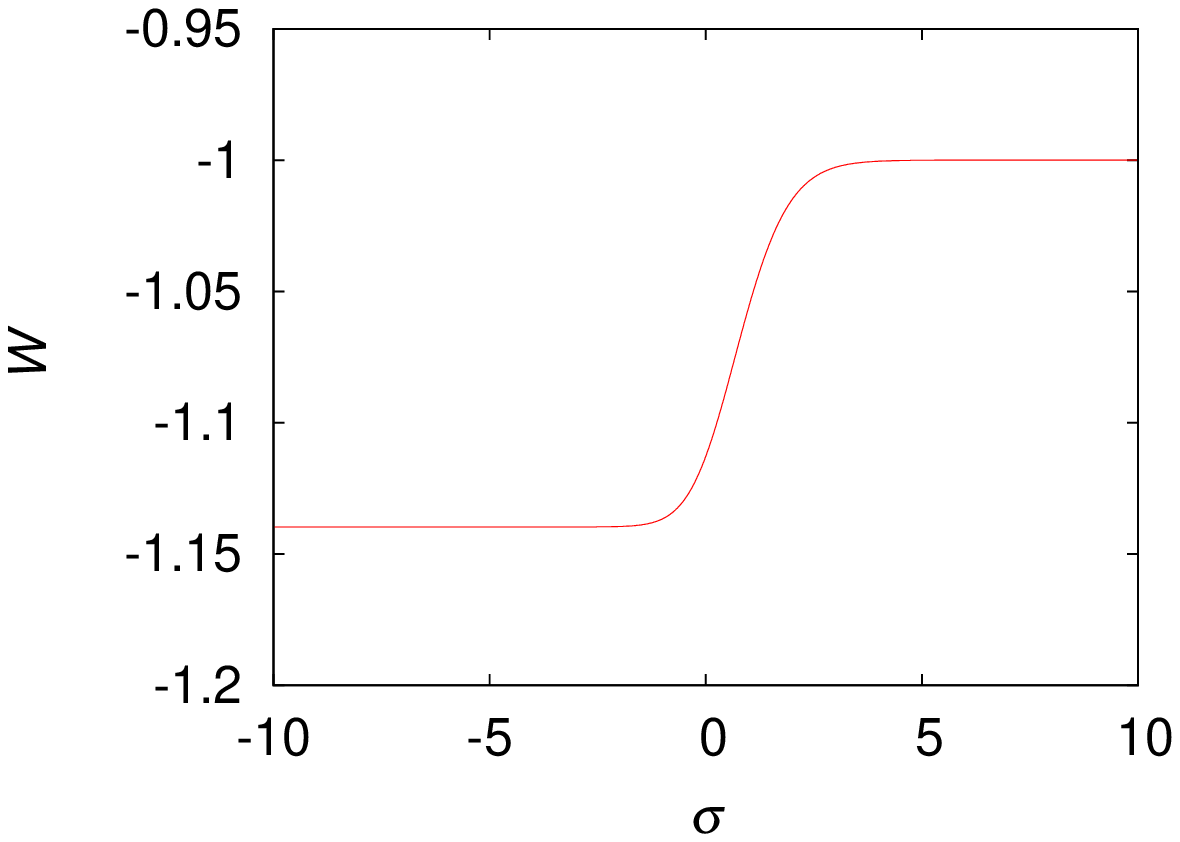} \\
(c)\includegraphics[width=0.47\textwidth]{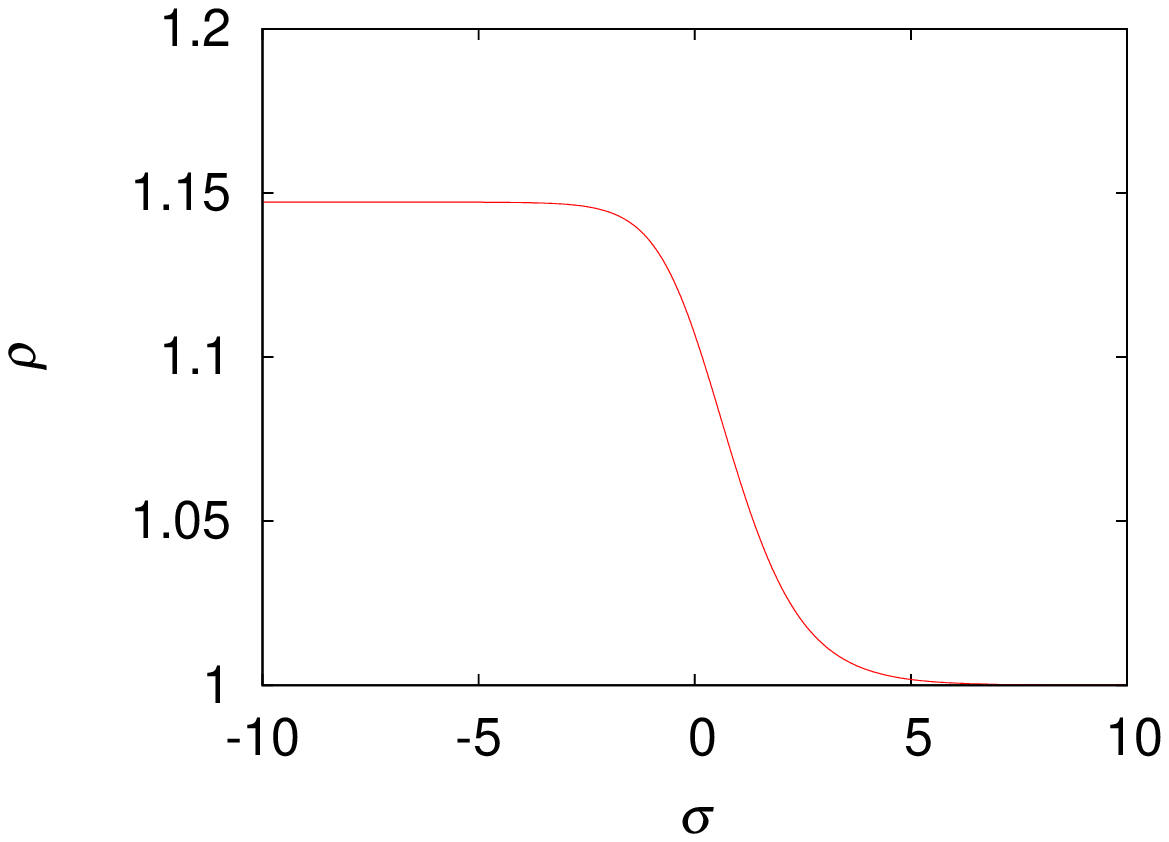} \hfill
(d)\includegraphics[width=0.47\textwidth]{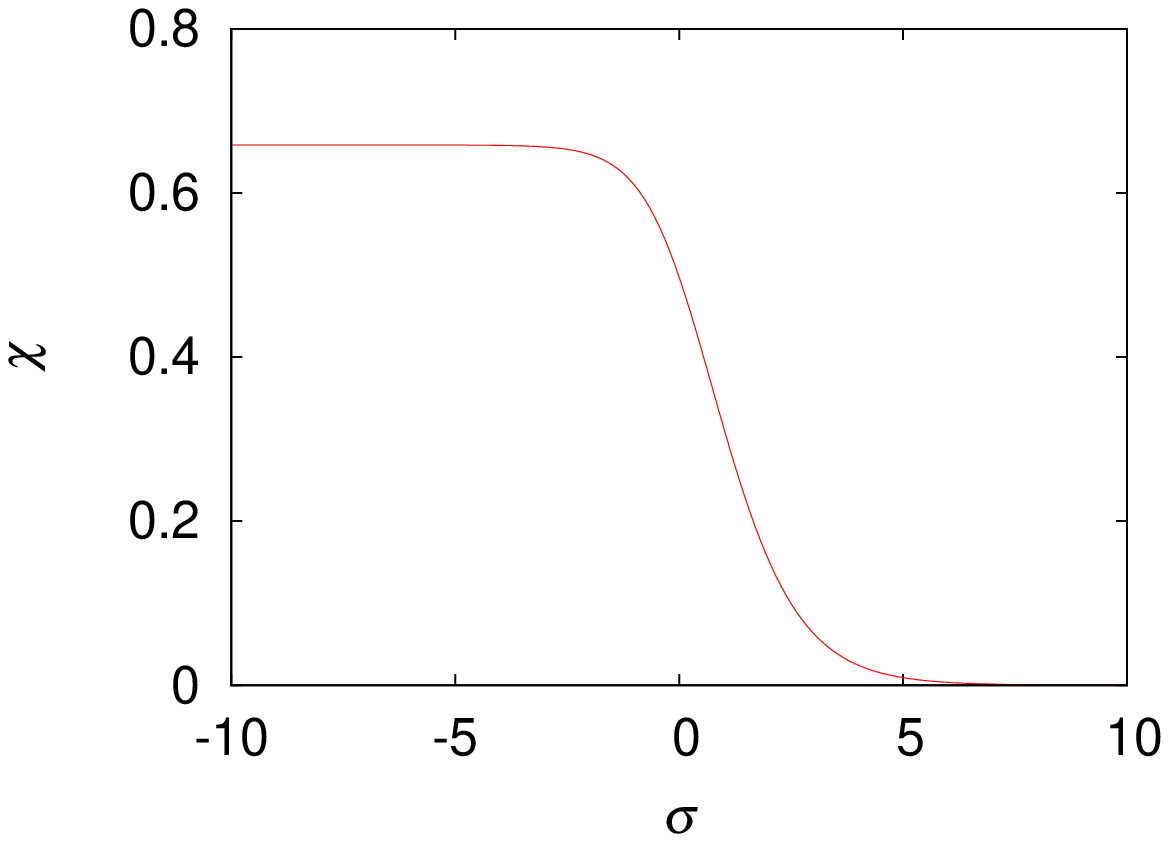}
\caption{(a) Contour plot of $W(\rho,\chi)$ with the solution of \eqref{a4c3:bg} connecting the UV and IR fixed points. (b) Plot of $W$ vs.\ $\sigma$ along the flow. (c,d) Plots of the scalar fields vs.\ $\sigma$. \label{a4c3:bg.fig}}
\end{figure}

We wish to study the case of the $SU(3)\times U(1)$ invariant RG flow in AdS$_4$/CFT$_3$. The $\mathcal{N}=2$ domain wall solution in $\mathcal{N}=8$, 4-dimensional gauged SUGRA was found in \cite{Ahn:2000aq,Ahn:2000mf} and lifted to M-theory in \cite{Corrado:2001nv}. 

The fake SUGRA system describing the flow contains two scalar fields, $\rho$ and $\chi$, with the sigma model metric
\begin{equation}
 \label{a4c3:G.metric}
  G_{\rho\rho}= 12 \rho^{-2}~,\quad G_{\chi\chi}=1~,\quad G_{\chi\rho}=0
\end{equation}
and the superpotential\footnote{We have adopted the expressions of \cite{Ahn:2000aq,Ahn:2000mf,Corrado:2001nv} to the conventions of Sec.~\ref{bdyn} and set the length scale of the asymptotic AdS region in the UV to unity.}
\begin{equation}
 \label{a4c3:W}
  W = -\frac3{8\rho^2} \left[ \cosh (2\chi) +1\right] 
      + \frac18\rho^6 \left[\cosh (2\chi) -3 \right]~.
\end{equation}
It will be helpful to introduce also a canonical basis $(\alpha,\chi)$ of scalars by defining
\begin{equation}
 \label{a4c3:alpha.def}
 \rho = \e{\frac{\alpha}{2\sqrt{3}}}~.
\end{equation}

The background equations \eqref{bdyn:BPS.2} read
\begin{equation}
 \label{a4c3:bg}
 \begin{split}
   \partial_\sigma \rho &= -\frac{\rho}{16W} \left[ \frac1{\rho^2}(\cosh (2\chi) +1) 
	+ \rho^6 (\cosh (2\chi) -3) \right]~,\\
   \partial_\sigma \chi &= -\frac1{4W} \left(\rho^6-\frac{3}{\rho^2}\right) \sinh (2\chi)~.
 \end{split}
\end{equation}
The fixed points we are interested in are the (attractive) UV fixed point at $\rho=1$ ($\alpha=0$), $\chi=0$ and the IR fixed point at $\rho^8=3$, $\cosh (2\chi)=2$. (Notice that $\rho$ is assumed positive, and amongst the two equivalent IR fixed points we pick the one with positive $\chi$.) The solution of \eqref{a4c3:bg}, which interpolates between the fixed points, can be found numerically and is illustrated in Fig.~\ref{a4c3:bg.fig}. For numerical stability, one sets the initial conditions for the field values very close to the IR fixed point, and the dynamics of \eqref{a4c3:bg} lets the solution flow to the UV fixed point with increasing $\sigma$. This numerical solution will serve as the domain wall background in the numerical treatment of the fluctuations, which follows.

Before continuing, we would like to make some comments on the integration constants for \eqref{a4c3:bg}.\footnote{A similar discussion for the flow in AdS$_5$/CFT$_4$ can be found in \cite{Freedman:1999gp} and in Sec.~\ref{a5c4:bg.sec}.} For large values of $\sigma$, \ie in the UV region, the scalar fields approach their respective UV fixed point values as
\begin{equation}
\label{a4c3:UVasymp}
  \text{large $\sigma$}:\qquad 
  \alpha(\sigma) \approx \hat{\alpha} \e{-\sigma}~,\quad 
  \chi(\sigma) \approx \hat{\chi} \e{-\sigma}~,
\end{equation}
with two coefficients $\hat{\alpha}$ and $\hat{\chi}$. However, only their ratio, or equivalently, an angle in field space, determines the direction of the flow. For the flow starting at the IR fixed point, the numerical solution yields
\begin{equation}
\label{a4c3:UVinvar}
\frac{\hat{\alpha}}{\hat{\chi}}\approx 0.6529~.
\end{equation}
A possible common factor can be absorbed by a shift of the radial variable $\sigma\to \sigma'=\sigma+\delta \sigma$.

Let us now consider the dimensions of the dual operators at the fixed points. This was done in \cite{Ahn:2000aq}, but we shall rederive them here for completeness. The dimensions of the operators dual to the scalars $\alpha$ and $\chi$ are determined by expanding $W$ about a fixed point to quadratic order. At the UV fixed point, one finds
\begin{equation}
 \label{a4c3:W.exp.UV}
  W = -\left( 1+\frac12 \alpha^2 +\frac12 \chi^2 \right) +\cdots~.
\end{equation}
Comparing this to the generic formula \eqref{asymp:W.exp}, we find that the scalars $\alpha$ and $\chi$ correspond to a doublet of relevant operators of dimension $\Delta=2$.

For the IR fixed point we introduce
\begin{equation}
 \label{a4c3:IR.vars}
  \alpha = \alpha_\IR + \talpha~, \quad \chi = \chi_\IR +\tchi~,
\end{equation}
where $\alpha_\IR$ and $\chi_\IR$ are the respective field values at the fixed point. Expanding $W$ to quadratic order yields
\begin{equation}
 \label{a4c3:W.exp.IR}
  W = -\frac1{L_\IR} \left( 1 +\frac12 \talpha^2 -2 \talpha \tchi \right) +\cdots~,
\end{equation}
with $L_\IR=2\cdot 3^{-3/4}$. Rotating the fields by
\begin{equation}
 \label{a4c3:IR.var.rot}
  \begin{pmatrix} \talpha' \\ \tchi' \end{pmatrix} = 
  \begin{pmatrix} \cos \varphi & -\sin\varphi \\ \sin\varphi& \cos\varphi \end{pmatrix}
  \begin{pmatrix} \talpha \\ \tchi \end{pmatrix}~,
\end{equation}
where
\begin{equation}
 \label{a4c3:IR.rot.angle}
  \cos \varphi = \left[ \frac12 \left( 1+ \frac1{\sqrt{17}} \right) \right]^{1/2}~, \quad
  \sin \varphi = \left[ \frac12 \left( 1- \frac1{\sqrt{17}} \right) \right]^{1/2}~,
\end{equation}
one brings the quadratic terms in \eqref{a4c3:W.exp.IR} into diagonal form and obtains
\begin{equation}
 \label{a4c3:IR.lambda}
  \lambda_{\talpha'} = 1 -\frac12 \sqrt{17}~, \quad 
  \lambda_{\tchi'} = 1 +\frac12 \sqrt{17}~.
\end{equation}
Hence, the dual operators $\op_{\talpha'}$ and $\op_{\tchi'}$ have dimensions
\begin{equation}
 \label{a4c3:IR.dims}
  \Delta_{\talpha'} = \frac12 \left( 1+ \sqrt{17}\right)~, \quad
  \Delta_{\tchi'} = \frac12 \left( 5+ \sqrt{17}\right)~.
\end{equation}
From the monotonicity relation $\partial_r W = W_a W^a \geq0$, which stems from \eqref{bdyn:BPS}, follows that the flow approaches the IR fixed point along the $\tchi'$ direction, and the irrelevant operator $\op_{\tchi'}$ controls the RG flow in the field theory \cite{Freedman:1999gp}.

\subsection{Spectral Functions}
\label{a4c3:spec.fun}

Here, we present our numerical results for the spectral functions of the two-point correlators for the doublet of operators $\op_\alpha$ and $\op_\chi$, which have UV conformal dimensions $\Delta_\alpha=\Delta_\chi=2$. As shown above, the renormalization group flow lifts the degeneracy of the dimensions and ends at the IR fixed point with two operators $\op_{\talpha'}$ and $\op_{\tchi'}$, the dimensions of which are given by \eqref{a4c3:IR.dims}. From the eigenvalues of the spectral function matrix, we shall be able to identify clearly the cross-over from the UV to the IR. Furthermore, we shall look for evidence of UV oscillations in the spectral functions, as in the toy model, but find no evidence of them.

The system of field equations \eqref{num:eom}, which we have to solve numerically, contains in $\fa$ the linearized fluctuations of the scalars $\alpha$ and $\chi$. As the kinetic term for these scalars is canonical, we have simply $D_\sigma=\partial_\sigma$. Furthermore, the matrix $M$ takes the form
\begin{equation}
   \label{a4c3:M}
   M = \frac1{W^2} 
     \begin{pmatrix}
        \frac{\rho^4}4 [\cosh (2\chi) -3][\cosh (2\chi)+1] & 
        \frac{\sqrt{3}}2 \rho^4 \sinh (2\chi) \\
        \frac{\sqrt{3}}2 \rho^4 \sinh (2\chi) &
        \frac1{16\rho^4} (\rho^8-3)\left[ 3(\rho^8+1) \cosh (2\chi) -\rho^8 +3 \right]
     \end{pmatrix}~,
\end{equation}
where the superpotential $W$ is given by \eqref{a4c3:W}, $\rho=\e{\alpha/(2\sqrt{3})}$, and the fields are evaluated on the ($\sigma$-dependent) background. As explained in Sec.~\ref{num:strat}, we use \eqref{num:eom.IR} in the IR region for numerical stability choosing $\sigma_0=0$.

\begin{figure}[th]
(a)\includegraphics[width=0.47\textwidth]{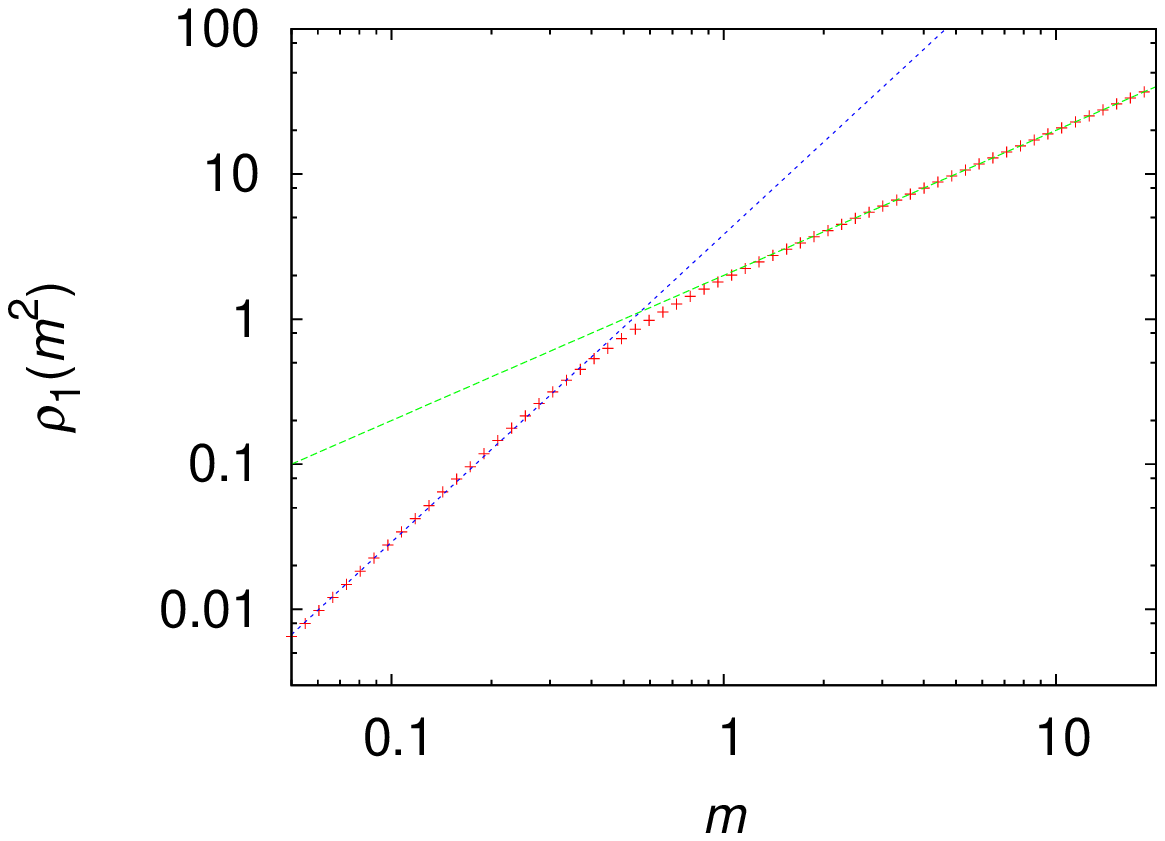} \hfill
(b)\includegraphics[width=0.47\textwidth]{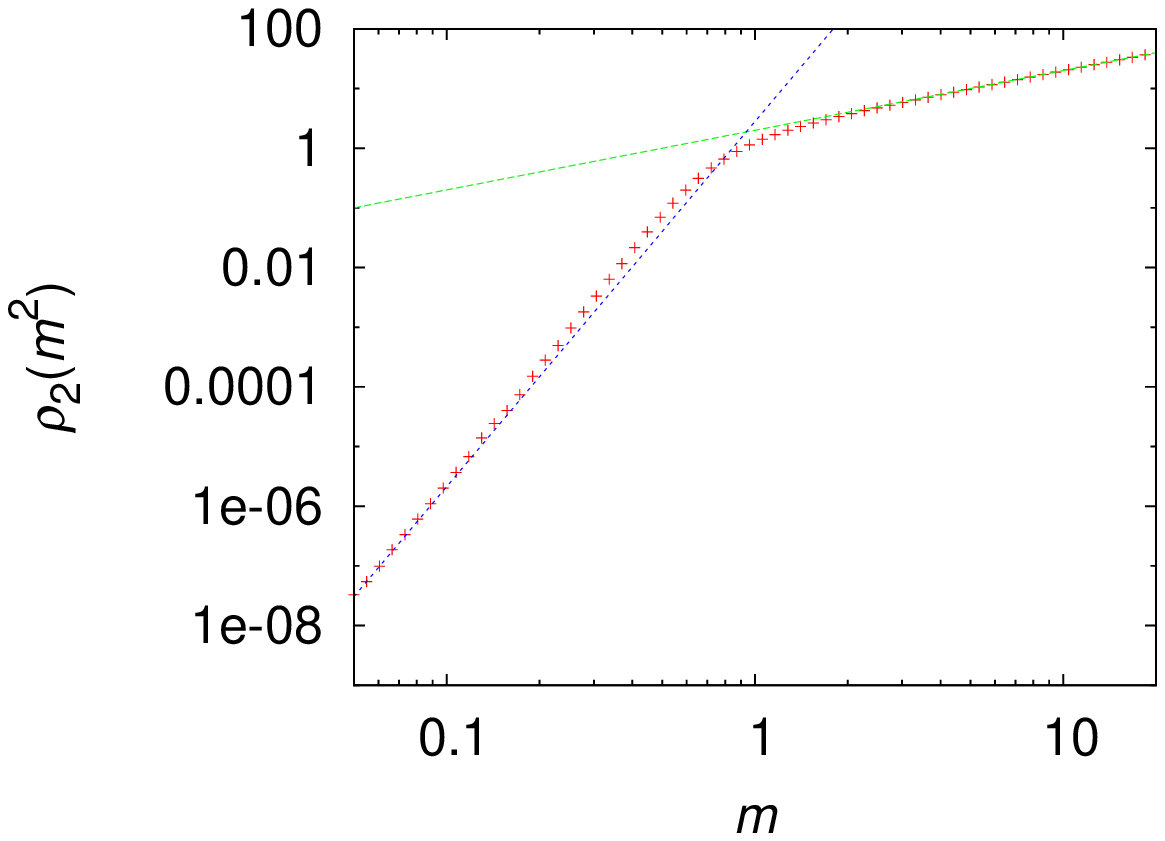}
\caption{Log-log plots of the spectral function eigenvalues $\rho_1$ (a) and $\rho_2$ (b) vs.\ $m$. The straight lines correspond to the expected IR and UV behaviours. The UV lines are given by $\rho=2m$, while the IR lines are (a) $\rho_1=3.83\, m^{2.12}$ and (b) $\rho_2=2.8\, m^{6.12}$.
\label{a4c3:spec.fig}}
\end{figure}

The results for the eigenvalues of the spectral function matrix are shown in Fig.~\ref{a4c3:spec.fig} and exhibit the following features. For large $m$, both eigenvalues show the expected UV behaviour \eqref{asymp:rho} with $\lambda=1/2$, \ie $\rho=2m$, without any sign of oscillations. (We have checked also for higher values of $m$ than those shown in the figure.) For small $m$, a fit of the log of the larger eigenvalue vs.\ $\ln m$ yields the formula $\ln \rho_1\approx 2.12 \ln m +1.34$, and the coefficient of $\ln m$ agrees nicely with the expected value for the operator $\op_{\talpha'}$,
\begin{equation}
   \label{a4c3:rho1.exp}
   2 \Delta_{\talpha'} -3 = \sqrt{17}-2 \approx 2.12~.
\end{equation}
A similar fit for the smaller eigenvalue in the IR region yields $\ln \rho_2 \approx 6.17 \ln m +1.21$. The slight disagreement of the slope from the expected value, which is $6.12$, is due to the larger numerical error for $\rho_2$.\footnote{The smaller eigenvalue $\rho_2$, which is several orders of magnitude smaller than $\rho_1$ in the region of interest, is calculated as the difference of two numbers of the order of $\rho_1$.} 

The results for the spectral functions indicate that the deformed CFT exhibits a simple cross-over behaviour under the RG flow, with a cross-over point  $m=\Lambda_c$. For $m>\Lambda_c$, its dynamics is governed by the undeformed $\mathcal{N}=8$ CFT, whereas for $m<\Lambda_c$ the IR conformal phase takes over. The numerical value of $\Lambda_c$ is subject to shifts of the radial variable $\sigma\to \sigma'=\sigma +\delta\sigma$, which just amounts to choosing a momentum scale.

\section{$SU(2) \times U(1)$ RG flow in $d=4$}
\label{a5c4}
\subsection{Fixed Points and Background Solution}
\label{a5c4:bg.sec}

This RG flow was discussed extensivley in \cite{Freedman:1999gp}. Here, we will give a brief review for completeness.

The fake SUGRA system describing the RG flow contains two scalar fields, $\rho$ and $\chi$, with the sigma model metric
\begin{equation}
 \label{a5c4:G.metric}
  G_{\rho\rho}= 6 \rho^{-2}~,\quad G_{\chi\chi}=1~,\quad G_{\chi\rho}=0
\end{equation}
and the superpotential 
\begin{equation}
 \label{a5c4:W}
  W = -\frac1{2\rho^2} \left[ \cosh (2\chi) +1\right] 
      + \frac14\rho^4 \left[\cosh (2\chi) -3 \right]~.
\end{equation}

A canonical basis of scalars $\beta, \chi$ is achieved by introducing\footnote{In the notation of \cite{Freedman:1999gp}, $\beta=\varphi_3$, $\chi=\varphi_1$.}
\begin{equation}
 \label{a5c4:alpha.def}
 \rho = \e{\frac{\beta}{\sqrt{6}}}~.
\end{equation}

\begin{figure}[th]
(a)\includegraphics[width=0.47\textwidth]{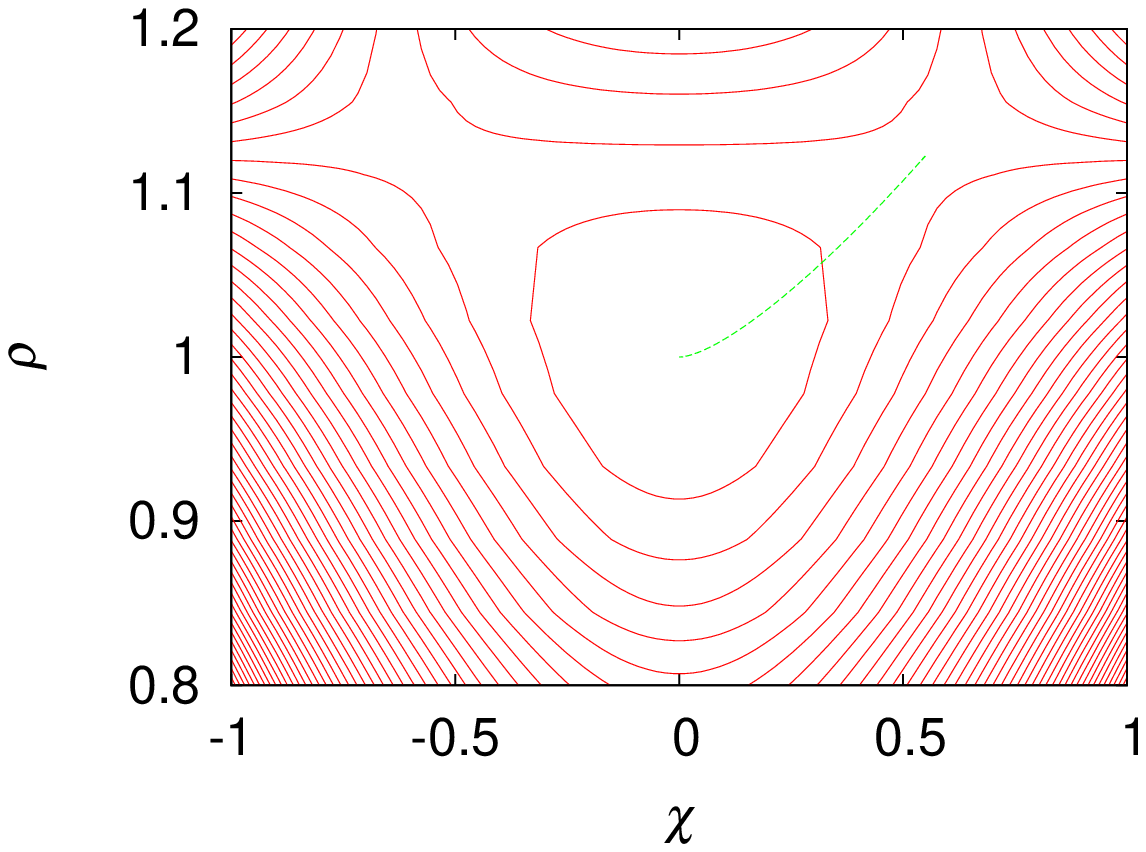} \hfill
(b)\includegraphics[width=0.47\textwidth]{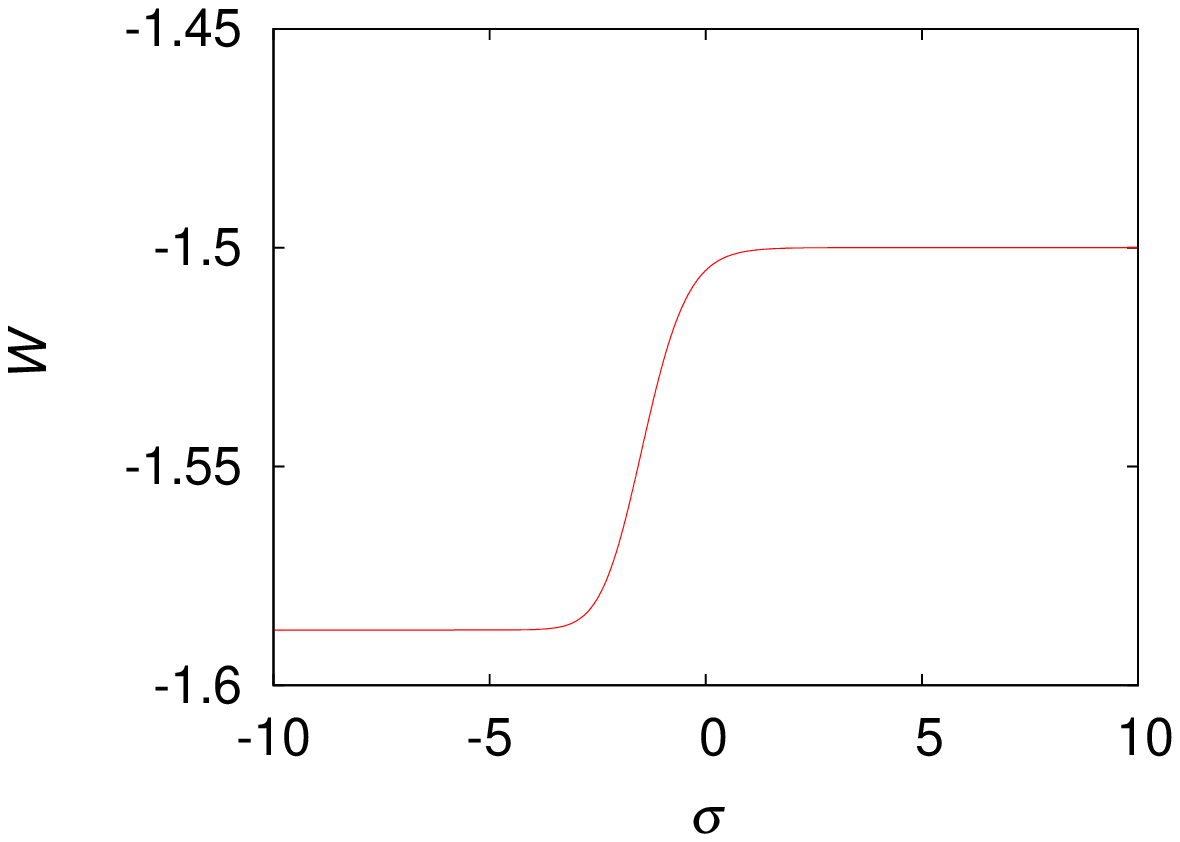} \\
(c)\includegraphics[width=0.47\textwidth]{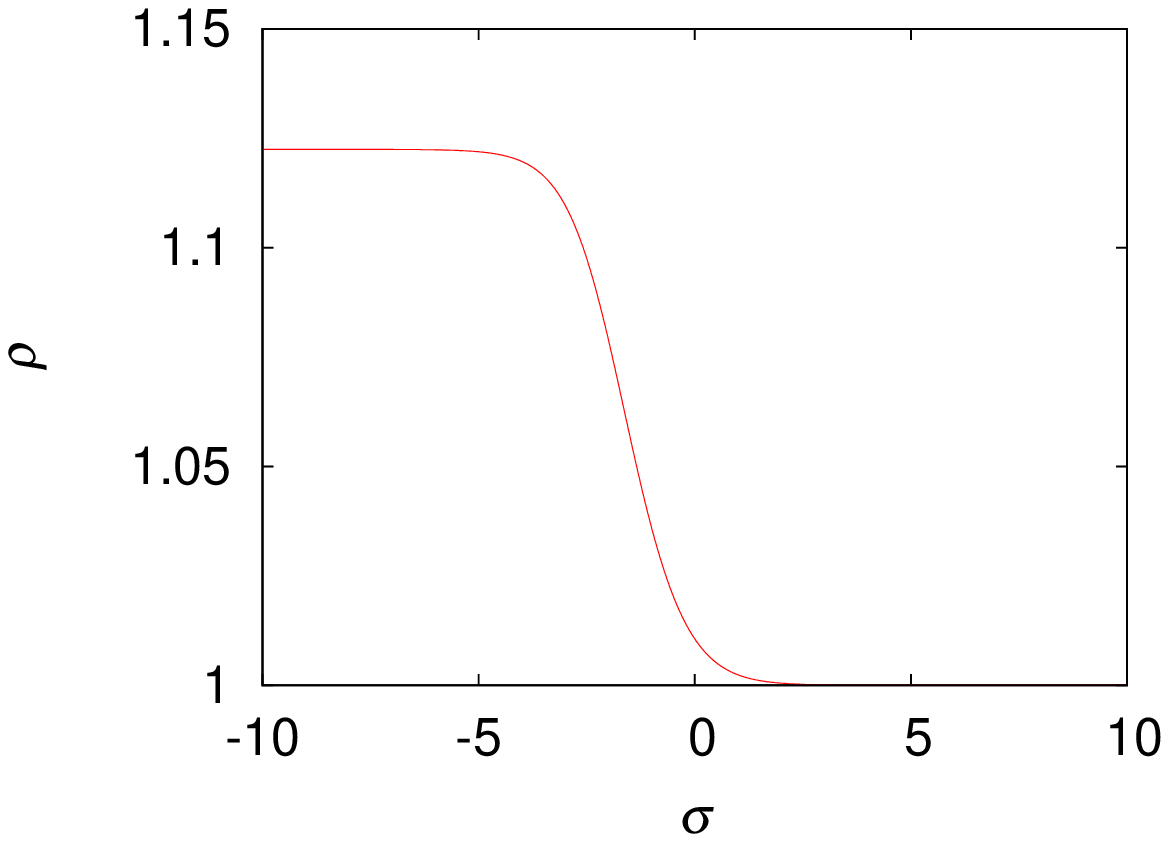} \hfill
(d)\includegraphics[width=0.47\textwidth]{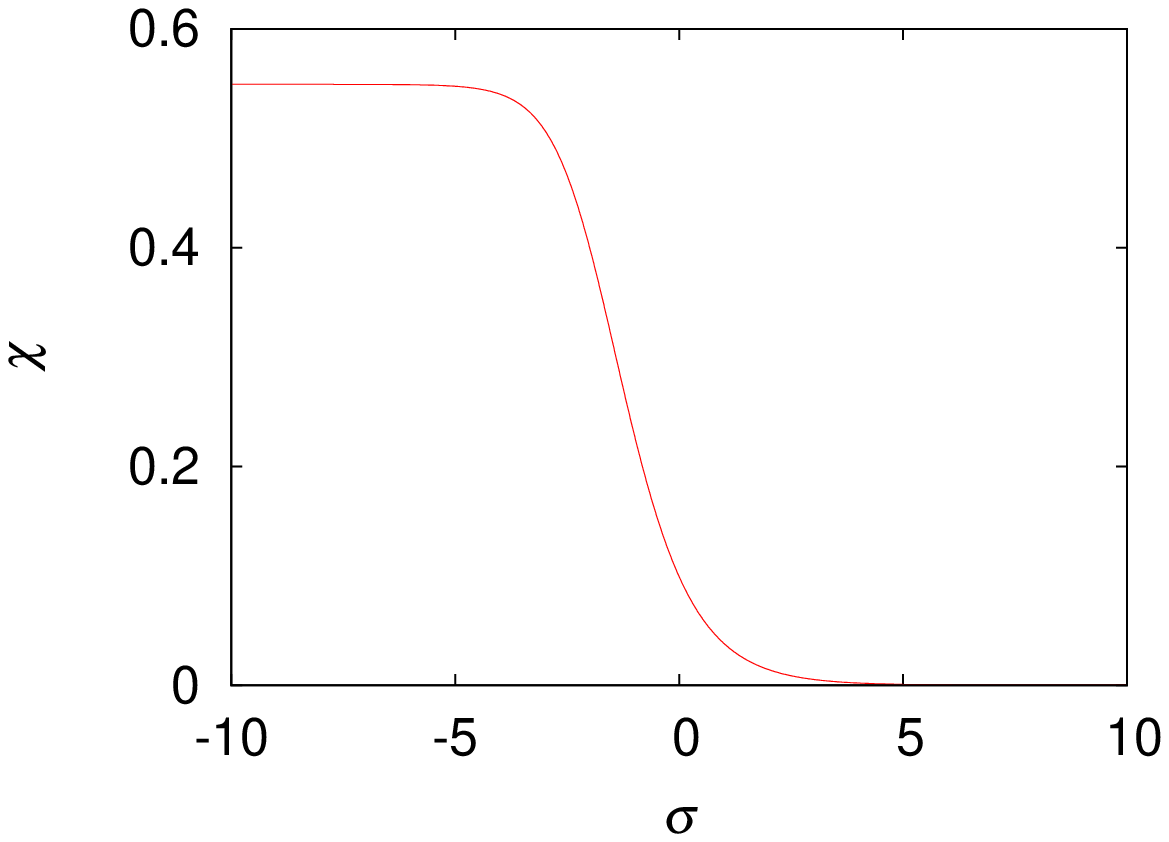}
\caption{(a) Contour plot of $W(\rho,\chi)$ with the solution of \eqref{a5c4:bg} connecting the UV and IR fixed points. (b) Plot of $W$ vs.\ $\sigma$ along the flow. (c,d) Plots of the scalar fields vs.\ $\sigma$. \label{a5c4:bg.fig}}
\end{figure}

The background equations \eqref{bdyn:BPS.2} read
\begin{equation}
 \label{a5c4:bg}
 \begin{split}
   \partial_\sigma \rho &= -\frac{\rho}{4W} \left[ \frac1{\rho^2}(\cosh (2\chi) +1) 
	+ \rho^4 (\cosh (2\chi) -3) \right]~,\\
   \partial_\sigma \chi &= -\frac3{4W} \left(\rho^4-\frac{2}{\rho^2}\right) \sinh (2\chi)~.
 \end{split}
\end{equation}
The fixed points we are interested in are the (attractive) UV fixed point at $\rho=1$ ($\beta=0$), $\chi=0$ and the IR fixed point at $\rho^6=2$, $\cosh (2\chi)=5/3$. The numerical solution of \eqref{a5c4:bg}, which interpolates between the fixed points, is illustrated in Fig.~\ref{a5c4:bg.fig}. This numerical solution will serve as the domain wall background in the numerical treatment of the fluctuations, which follows.

The background behaviour of the fields near the UV fixed point is
\begin{equation}
\label{a5c4:UVasymp}
  \text{large $\sigma$}:\qquad 
  \chi(r) \approx \hat{\chi} \e{-\sigma}~, \quad
  \beta(r) \approx \hat{\beta} \e{-2\sigma} +\sqrt{\frac83} \hat{\chi}^2 \sigma \e{-2\sigma}~,
\end{equation}
with two coefficients $\hat{\beta}$ and $\hat{\chi}$. The combination of them, which is invariant under a shift of the radial variable, $\sigma\to \sigma'=\sigma+\delta \sigma$, is
\begin{equation}
\label{a5c4:UVinvar}
   \frac{\hat{\beta}}{\hat{\chi}^2} +\sqrt{\frac83} \ln \hat{\chi} \approx -1.469(6)~,
\end{equation}
which agrees with the result of \cite{Freedman:1999gp}.

Let us rederive the dimensions of the dual operators at the fixed points. At the UV fixed point, one finds
\begin{equation}
 \label{a5c4:W.exp.UV}
  W = -\left( \frac32+ \beta^2 +\frac12 \chi^2 \right) +\cdots~.
\end{equation}
Comparing this to the generic formula \eqref{asymp:W.exp}, we find that the scalars $\beta$ and $\chi$ correspond to relevant operators of dimensions $\Delta_\beta=2$ and $\Delta_\chi=3$, respectively.

For the IR fixed point we introduce
\begin{equation}
 \label{a5c4:IR.vars}
  \beta = \beta_\IR + \tbeta~, \quad \chi = \chi_\IR +\tchi~,
\end{equation}
where $\beta_\IR$ and $\chi_\IR$ are the respective field values at the fixed point. Expanding $W$ to quadratic order yields
\begin{equation}
 \label{a5c4:W.exp.IR}
  W = \frac1{L_\IR} \left( -\frac32 +\sqrt{6} \tbeta \tchi -\tbeta^2 \right) +\cdots~,
\end{equation}
with $L_\IR=3\cdot 2^{-5/3}$. Rotating the fields by
\begin{equation}
 \label{a5c4:IR.var.rot}
  \begin{pmatrix} \tbeta' \\ \tchi' \end{pmatrix} = 
  \begin{pmatrix} \cos \varphi & -\sin\varphi \\ \sin\varphi& \cos\varphi \end{pmatrix}
  \begin{pmatrix} \tbeta \\ \tchi \end{pmatrix}~,
\end{equation}
where
\begin{equation}
 \label{a5c4:IR.rot.angle}
  \cos \varphi = \left[ \frac12 \left( 1+ \frac1{\sqrt{7}} \right) \right]^{1/2}~, \quad
  \sin \varphi = \left[ \frac12 \left( 1- \frac1{\sqrt{7}} \right) \right]^{1/2}~,
\end{equation}
one brings the quadratic terms in \eqref{a5c4:W.exp.IR} into diagonal form and obtains
\begin{equation}
 \label{a5c4:IR.lambda}
  \lambda_{\tbeta'} = 1 -\sqrt{7}~, \quad 
  \lambda_{\tchi'} = 1 + \sqrt{7}~.
\end{equation}
Hence, the dual operators $\op_{\tbeta'}$ and $\op_{\tchi'}$ have dimensions
\begin{equation}
 \label{a5c4:IR.dims}
  \Delta_{\tbeta'} = 1+ \sqrt{7}~, \quad
  \Delta_{\tchi'} = 3+ \sqrt{7}~.
\end{equation}
The background approaches the IR fixed point along the $\tchi'$ direction, and the irrelevant operator $\op_{\tchi'}$ controls the RG flow in the field theory.

\subsection{Spectral Functions}
\label{a5c4:spec.fun}

Here, we present our numerical results for the spectral functions of the two-point correlators for the operators $\op_\beta$ and $\op_\chi$, which have UV conformal dimensions $\Delta_\beta=2$ and $\Delta_\chi=3$, respectively. As shown above, the renormalization group flow ends at the IR fixed point with two operators $\op_{\tbeta'}$ and $\op_{\tchi'}$, the dimensions of which are given by \eqref{a5c4:IR.dims}. As in the AdS$_4$/CFT$_3$ case, the spectra show a cross-over behaviour along the flow and no signs of oscillations.

The system of field equations \eqref{num:eom}, which we have to solve numerically, contains in $\fa$ the linearized fluctuations of the scalars $\beta$ and $\chi$. As the kinetic term for these scalars is canonical, we have simply $D_\sigma=\partial_\sigma$. Furthermore, the matrix $M$ takes the form
\begin{equation}
   \label{a5c4:M}
   M = \frac3{2W^2} 
     \begin{pmatrix}
        \frac{3\rho^2}4 [\cosh (2\chi) -3][\cosh (2\chi)+1] & 
        \sqrt{6} \rho^2 \sinh (2\chi) \\
        \sqrt{6} \rho^2 \sinh (2\chi) &
        \frac1{4\rho^4} (\rho^6-2)\left[ (3\rho^6+2) \cosh(2\chi) -\rho^6 +2 \right]
     \end{pmatrix}~,
\end{equation}
where the superpotential $W$ is given by \eqref{a5c4:W}, $\rho=\e{\beta/\sqrt{6}}$, and the fields are evaluated on the ($\sigma$-dependent) background.

\begin{figure}[t]
(a)\includegraphics[width=0.47\textwidth]{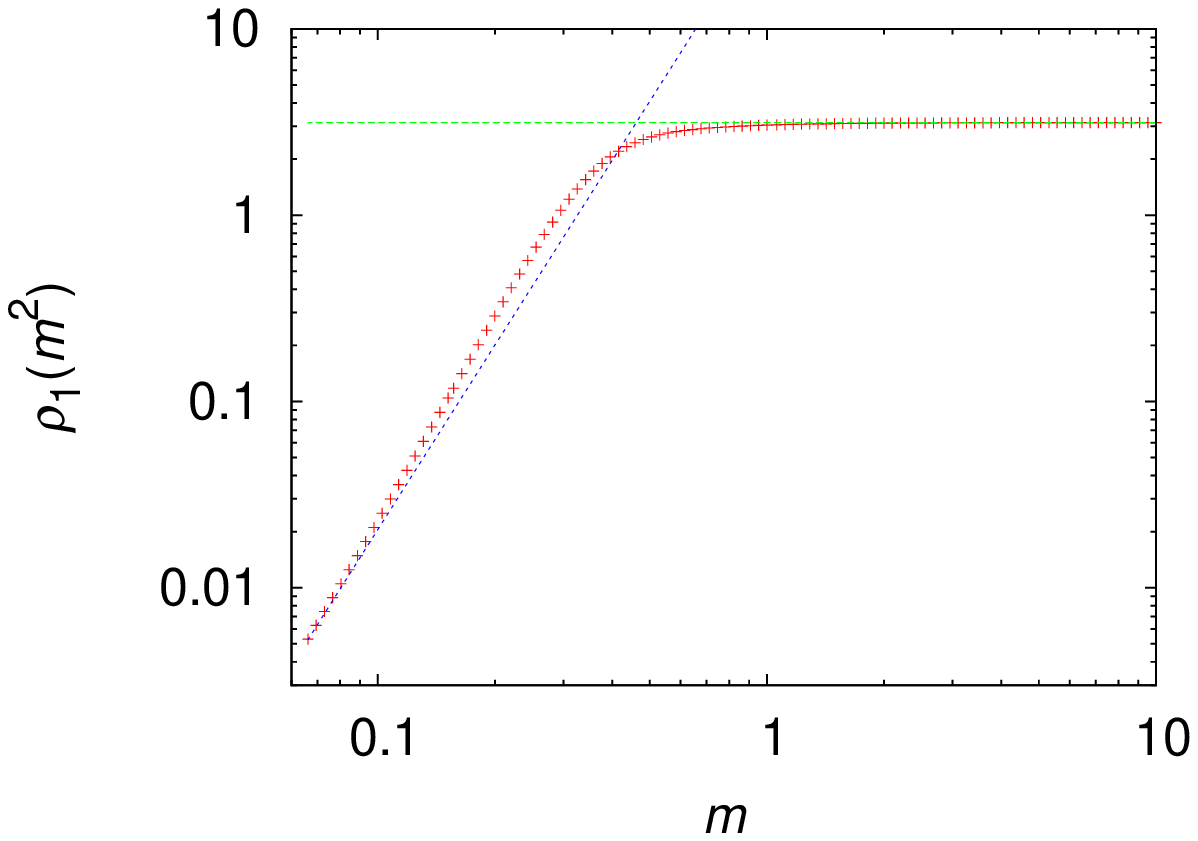} \hfill
(b)\includegraphics[width=0.47\textwidth]{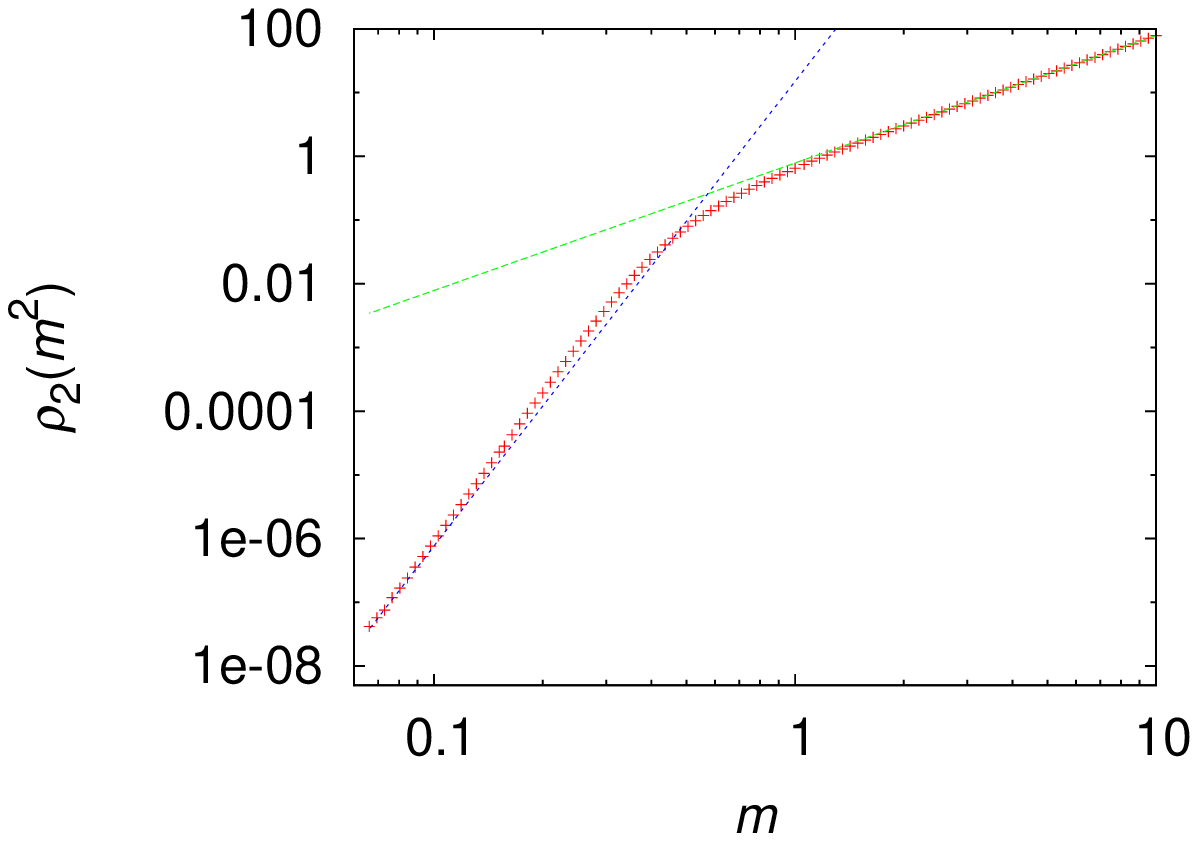}
\caption{(a) and (b) log-log plots of the spectral function eigenvalues $\rho_1$ and $\rho_2$ vs.\ $m$, respectively. The straight lines correspond to the expected IR and UV behaviours. The UV lines follow from \eqref{asymp:rho} with (a) $\lambda=0$ and (b) $\lambda=1$. The IR lines are (a) $\rho_1=40\, m^{3.29}$ and (b) $\rho_2=15\, m^{7.29}$. \label{a5c4:spec.fig}}
\end{figure}

The results for the eigenvalues of the spectral function matrix are shown in Fig.~\ref{a5c4:spec.fig} and exhibit a clear cross-over behaviour from the UV to the IR. For large $m$, the spectral functions exhibit the expected UV behaviour \eqref{asymp:rho} for operators of dimensions $\Delta_\beta=2$ and $\Delta_\chi=3$,
\begin{equation}
   \label{a5c4:UV.rho}
	\rho_1 = \rho_\beta \approx \pi~,\qquad \rho_2 = \rho_\chi = \frac{\pi}4 m^2~,
\end{equation}
respectively. Again, there is no sign of oscillations.

For small $m$, a fit of $\ln \rho$ vs.\ $\ln m$ yields the relations $\rho_1\approx  3.5\ln m+4.3$ and $\rho_2\approx 7.6 \ln m+3.5$, while the values expected for the slopes are, respectively,
\begin{equation}
   \label{a5c4:rho.exp}
   2 (\sqrt{7} -1) \approx 3.29~,\qquad 2 (\sqrt{7} +1) \approx 7.29~,
\end{equation}
as one obtains from the IR dimensions \eqref{a5c4:IR.dims}.
We suspect that the slight disagreement stems from numerical issues, but this question deserves further investigation. We just note that, with a numerical accuracy of the integration routine of about $10^{-6}$, it is impossible to calculate the eigenvalues for lower $m$. At $m=0.08$, the eigenvalues differ already by five orders of magnitude. 

\section*{Acknowledgments}

I would like to thank J.~Erdmenger and M.~Haack for stimulating discussions. This work has been supported in part by the European Community's Human Potential Programme under
contract MRTN-CT-2004-005104 'Constituents, fundamental forces and symmetries of the 
universe' and by the Italian Ministry of Education and Research (MIUR), project 2005-023102.

\bibliographystyle{JHEP}
\bibliography{flow}
\end{document}